\documentclass[12pt,a4,portrait]{article}

\usepackage{mathtools}
\usepackage{tocbibind}

\usepackage{relsize}
\usepackage{mathenv}
\usepackage[compress]{cite}
\usepackage{booktabs}
\usepackage{graphicx}
\usepackage[toc,page]{appendix}
\usepackage[utf8]{inputenc}
\usepackage[T1]{fontenc}
\usepackage{xcolor}
\usepackage[ocgcolorlinks,colorlinks=true,linkcolor=blue,citecolor=blue,urlcolor=blue]{hyperref}

\begin{document}


	


\title{{\bf Invariant slow-roll parameters in scalar-tensor theories}}

\author{Piret Kuusk\thanks{piret.kuusk@ut.ee}, Mihkel R\"unkla\thanks{mrynkla@ut.ee}, Margus Saal\thanks{margus.saal@ut.ee} and Ott Vilson\thanks{ovilson@ut.ee}\\
	{\normalsize Institute of Physics, University of Tartu, W. Ostwaldi Str 1, 50411, Tartu, Estonia}}

\maketitle


\begin{abstract}
A general scalar-tensor theory can be formulated in different parametrizations that are related by a conformal rescaling of the metric and a scalar field redefinition. We compare formulations of slow-roll regimes in the Einstein and Jordan frames using quantities that are invariant under the conformal rescaling of the metric and transform as scalar functions under the reparametrization of the scalar field. By comparing spectral indices, calculated up to second order, we find that the frames are equivalent up to this order, due to the underlying assumptions.
\vspace{0.2cm}
\end{abstract}




\section{Introduction}

The phase of accelerated expansion in the early universe called inflationary epoch was originally motivated to solve the problems of homogeneity and isotropy, as well as to explain why one does not observe any cosmic relics \cite{Guth:1980zm,Linde:1981mu,Linde}. The epoch of accelerated expansion can be realized by a scalar field minimally coupled to gravity in the framework of general relativity, provided that the potential of the scalar field dominates over its kinetic energy
\cite{Steinhardt:1984,Mukhanov:2005sc,Lyth:2009zz}.
This is achieved in the so-called slow-roll regime, where one imposes slow-roll conditions for the scalar 
field or flatness conditions for the potential  \cite{Liddle:1994dx}. In the slow-roll regime the expansion rate 
is almost exponential and this can explain the nearly scale-invariant spectrum of quantum fluctuations of the 
scalar field produced during inflation \cite{Lyth:2009zz}.

The scenario of the slow-roll inflation with a minimally coupled scalar field can in general be conformed to observations. However, it has problems giving rise to criticism \cite{Ijjas:2014nta}, and the Planck mission measurements of cosmic microwave radiation disfavour the simplest model of inflation with a minimally coupled scalar field \cite{Ade:2015lrj}.
The Planck mission results \cite{Ade:2015lrj} seem to favour Starobinsky model \cite{Starobinsky:1980te}, non-minimal Higgs inflation \cite{Bezrukov:2007ep}, or generalized models with alpha-attractors \cite{Kallosh:2013hoa}, which all deal with non-minimal couplings. By 
introducing a non-minimal coupling one works in the framework of the scalar-tensor theory (STG) \cite{Barrow,Faraoni:1998qx,Capozziello:2010zz,Chiba:2014sva} rather than in general relativity with a minimally coupled scalar field. Many modified gravity theories \cite{Clifton:2011jh} can be recast into STG form, e.g.\ $f(R)$ \cite{Faraoni.Sotiriou}, scale-free \cite{Shaposhnikov:2008xi}, and non-local \cite{sasaki:nonlocal} theories.
 
Minimally coupled (MC) and non-minimally coupled (NMC) theories are in principle different, because the latter permits transformations which consist of rescaling of the metric and reparametrization of the scalar field. This allows one to perform  transformations between different conformal frames. For example, a NMC theory can be transformed to such a conformal frame, where the scalar field is minimally coupled  either to curvature (the Einstein frame) or to matter (the Jordan frame). In the case of the standard MC theory these frames coincide and hence it contains only one metric, which corresponds to the measurable one \cite{Faraoni:1998qx,Capozziello:2010zz,Dicke:1962gz}. In the case of NMC theories the choice of frame is arbitrary from the mathematical point of view, and one can work in either frame, but there is an ongoing debate on their physical equivalence \cite{physical.equivalence,Chiba:2013mha,Postma:2014vaa}.

Since inflation is driven by the scalar field, one can neglect the matter terms yielding that for inflationary dynamics, the NMC theory in Einstein frame and the MC theory have identical equations. 
This provides a method for generalizing the standard slow-roll regime from the MC theory to the general STG. However, in the MC theory the Einstein and Jordan frames simply coincide, and generalization of the results from the MC theory to the Einstein frame of the NMC theory must be taken with care. For instance, in the NMC theory one finds that the number of inflationary $\mathrm{e}$-folds in different frames is not equal \cite{Higgs.Inflation.and.Naturalness,vandeBruck:2015gjd} (but cf.\ \cite{Chiba:2014sva}). Further, in the MC theory the measurement of the amplitude of tensor fluctuations would determine the scale of inflation, whereas in the case of the NMC theory it would not completely determine the scale of inflation \cite{ArmendarizPicon:2015dma}.

Regarding the Jordan frame as physical motivates one to consider the slow-roll inflation directly in the Jordan frame. This can be achieved by defining the generalized or extended slow-roll conditions \cite{vandeBruck:2015gjd,Morris:2001ad,Chiba:2008ia}. In addition to the slow-roll parameters controlling the flatness of the effective potential, one has parameters for the non-minimal coupling. As pointed out in \cite{Torres:1996fr} and further discussed in \cite{Morris:2001ad}, if the slow-roll hierarchy is assumed for the scalar field, then its functions with sufficiently converging power series obey an analogous hierarchy. The calculations of the spectral indices have been carried out up to first order in slow-roll parameters and results in different frames agree \cite{Chiba:2008ia,Kaiser:1994vs,Kaiser:1995nv,Noh:2001ia}.

In this paper we compare the slow-roll regimes in the Einstein and Jordan frames using the formalism of invariant quantities \cite{JKSV.Invariants,Jarv:2015kga}. The main advantage of this formalism is the ability to quickly move between different parametrizations. We introduce scalar and tensorial invariants, the latter ones include the components of the invariant metric.
Note that the choice of the invariant metric is ambiguous and hence the definition of the invariant metric made e.g.\ in \cite{Postma:2014vaa} or \cite{Wetterich:2015ccd} is not unique. We would also like to address the discrepancy found in \cite{Nozari:2010uu} regarding the spectral indices up to second order in slow-roll parameters. Although our calculation is not fully general we find that under certain assumptions the spectral indices in both frames agree up to second order in slow-roll parameters. 

The outline of the paper is as follows. 
In section~\ref{Sec:General.theory} we present the action functional for the general non-minimally coupled scalar-tensor theory, and equations of motion for Friedmann-Lema\^itre-Robertson-Walker (FLRW) cosmological models in terms of scalar and tensorial quantities which are invariant under the conformal rescaling of the metric and transform as scalar functions under the reparametrization of the scalar field. 
In section~\ref{Sec:Slow.roll} we introduce slow-roll conditions in the invariant Einstein and Jordan frames and discuss their correspondence. In section~\ref{Sec:Spectral.indices} we consider the frame-(in)dependence of spectral indices up to second order in slow-roll parameters. 
In section~\ref{Sec:Summary} we conclude with a brief summary and outlook. 
Finally, in appendices~\ref{App:Invariants.in.different.parametrizations}~and~\ref{App:Slow.roll.parameters.in.different.parametrizations} we present invariants and slow-roll parameters in different parametrizations.


\section{General theory}\label{Sec:General.theory}


\subsection{General action functional}
\label{general.action.functional}


\subsubsection{Minimally coupled theory}

The action functional for gravity $g_{\mu\nu}$, a minimally coupled scalar field $\phi$ and matter fields $\chi$ can be written as
\begin{equation}
\label{nmcsf}
S = \frac{1}{16 \pi G_{\mathrm{N}}}\int_{V_4} \mathrm{d}^4x \sqrt{-g}
\left\lbrace R - 2 g^{\mu\nu} \nabla_\mu\phi \nabla_\nu\phi - 2\ell^{-2}{\mathcal V}(\phi) \right\rbrace
+ S_\mathrm{m} \left[ g_{\mu\nu},\chi \right] \,.
\end{equation}
Here $G_{\mathrm{N}}$ is Newton's gravitational constant, the determinant of the metric  $g_{\mu\nu}$ is $g$, whereas $R$ denotes the Ricci scalar with respect to the Levi-Civita connection $\nabla$, $\mathcal{V}(\phi)$ is the dimensionless scalar potential and $S_\mathrm{m}$ stands for the action functional for the matter fields $\chi$. The positive constant parameter $\ell$ has a dimension of length. For describing inflation in this framework one usually assumes $S_{\mathrm{m}} = 0$ and
the scalar field is in the role of matter. Leaving out the matter part of the action does not cause any problems, since the metric in the Ricci scalar and the metric in the action $S_{\mathrm{m}}$ are the same.


\subsubsection{Non-minimally coupled theory}

Let us consider the general action functional for a scalar-tensor theory of gravity \cite{Flanagan}
\begin{equation}
\label{fl.moju}
S = \frac{1}{2\kappa^2}\int_{V_4}\mathrm{d}^4x\sqrt{-g}\left\lbrace {\mathcal A}(\Phi)R-
{\mathcal B}(\Phi)g^{\mu\nu}\nabla_\mu\Phi \nabla_\nu\Phi - 2\ell^{-2}{\mathcal V}(\Phi)\right\rbrace 
+ S_\mathrm{m}\left[\mathrm{e}^{2\alpha(\Phi)}g_{\mu\nu},\chi\right] \,,
\end{equation}
which contains four arbitrary dimensionless functions of the dimensionless scalar field $\Phi$:
${\mathcal A}(\Phi)$, ${\mathcal B}(\Phi)$, ${\mathcal V}(\Phi)$ and
$\alpha(\Phi)$. The constant $\kappa^2$ has a dimension of Newton's constant $G_{\mathrm{N}}$.

By considering the conformal transformation of the metric and the scalar field redefinition
\begin{subequations}
%
\begin{eqnarray}
\label{conformal.transformation}
	g_{\mu\nu} &=& \mathrm{e}^{2\bar{\gamma}(\bar{\Phi})}\bar{g}_{\mu\nu} \,, \\
\label{field.redefinition}
	\Phi &=& \bar{f}(\bar{\Phi}) \,
\end{eqnarray}
%
\end{subequations}
one can fix two functional degrees of freedom out of the four functions $\{ \mathcal{A},\, \mathcal{B},\, \mathcal{V},\, \alpha \}$. Fixing any two functional degrees of freedom out of four in a particular manner is called choosing a parametrization. A conformal rescaling of the metric is referred to as a change of frame and a scalar field redefinition is called a reparametrization of the scalar field.

Under the transformations (\ref{conformal.transformation})-(\ref{field.redefinition}) the action functional (\ref{fl.moju}) preserves its structure up to the boundary term
if the functions of the scalar field transform accordingly \cite{Flanagan,Burns.et.al}:
\begin{subequations}
%
\begin{eqnarray}
\label{Flanagan.A}
	\bar{\mathcal{A}}(\bar{\Phi}) &=& \mathrm{e}^{2\bar{\gamma}(\bar{\Phi})}
	{\mathcal A} \left( {\bar f}( {\bar \Phi})\right) \,,\\
	\label{Flanagan.B}
	{\bar {\mathcal B}}({\bar \Phi}) &=& \mathrm{e}^{2{\bar \gamma}({\bar \Phi})}\left( 
	\left(\bar{f}^\prime\right)^2{\mathcal B}\left(\bar{f}(\bar{\Phi})\right) -
	 6\left(\bar{\gamma}^{\,\prime}\right)^2 {\mathcal A} \left(\bar{f}(\bar{\Phi})\right) -
	  6\bar{\gamma}^{\,\prime}\bar{f}^\prime \mathcal{A}^\prime \right) \,, \\
	\label{Flanagan.V}
	\bar{{\mathcal V}}(\bar{\Phi}) &=& \mathrm{e}^{4\bar{\gamma}(\bar{\Phi})} \, {\mathcal V}\left(\bar{f}(\bar{\Phi})\right) \,, \\
	\label{Flanagan.alpha}
	\bar{\alpha}(\bar{\Phi}) &=& \alpha\left(\bar{f}(\bar{\Phi})\right) + \bar{\gamma}(\bar{\Phi})\,.
\end{eqnarray}
%
\end{subequations}
Here the prime denotes derivative with respect to the argument,  e.g.\ for a barred quantity 
$\bar{\gamma}^{\,\prime}  \equiv  \mathrm{d}\bar{\gamma}(\bar{\Phi}) / \mathrm{d}\bar{\Phi}$, and for a quantity without a bar
$\mathcal{A}^\prime \equiv \mathrm{d} \mathcal{A}(\Phi)/\mathrm{d} \Phi$.  


\subsection{Invariants}
\label{Subsec:Invariant.quantities}


\subsubsection{Scalar invariants}
 
By a straightforward calculation one can verify that quantities \cite{JKSV.Invariants,Vilson.Some} 
\begin{subequations}
%
\begin{eqnarray}
\label{I.1.I.2}
\mathcal{I}_1(\Phi) &\equiv& \frac{\mathrm{e}^{2\alpha(\Phi)}}{\mathcal{A}(\Phi)} \,, \qquad \qquad
\mathcal{I}_2(\Phi) \equiv \frac{\mathcal{V}(\Phi)}{\left(\mathcal{A}(\Phi)\right)^2}\,, \\
\label{I.3}
\mathcal{I}_3(\Phi) &\equiv& \pm \int \left( \frac{2\mathcal{AB} + 3\left(\mathcal{A}^\prime\right)^2}{4\mathcal{A}^2} \right)^{\frac{1}{2}}\,\mathrm{d}\Phi   \equiv \pm \int \sqrt{\mathcal{F}(\Phi)}\,\mathrm{d}\Phi \,
\end{eqnarray}
%
\end{subequations}
are invariant under a local rescaling of the metric and transform as scalar functions under a redefinition of the scalar field due to transformation properties (\ref{Flanagan.A})-(\ref{Flanagan.alpha}) of the arbitrary functions
$\lbrace \mathcal{A}$, $\mathcal{B}$, $\mathcal{V}$, $\alpha \rbrace$. The numerical values of quantities (\ref{I.1.I.2})-(\ref{I.3}) in a spacetime point are invariant under transformations (\ref{conformal.transformation})-(\ref{field.redefinition}) and therefore we shall refer to these quantities as invariants \cite{JKSV.Invariants,Vilson.Some}.

The quantities (\ref{I.1.I.2})-(\ref{I.3}) allow us to characterize a theory under consideration in an invariant manner. Namely, if $\mathcal{I}_1$ is a dynamical function, then the theory possesses non-minimal coupling. A dynamical $\mathcal{I}_2$ describes non-trivial self-interactions of the scalar field, while $\mathcal{I}_2$ being (possibly vanishing) constant states that the equation of motion for the scalar field does not depend on the potential $\mathcal{V}$. The third invariant $\mathcal{I}_3$ can only be constant if the scalar field is not a propagating degree of freedom. The integrand $\sqrt{\mathcal{F}}\, \mathrm{d}\Phi$ can be interpreted as an invariant volume form in the $1$-dimensional space of the scalar field. This fact reveals itself more evidently if a theory with several scalar fields is considered \cite{JKV.2016}. Hence one can also consider $\mathcal{I}_3$ as an invariant length in the $1$-dimensional space of the scalar field (cf. footnote 3 in \cite{Flanagan}).

In addition to the three basic invariants (\ref{I.1.I.2})-(\ref{I.3}) with particular meaning one can construct infinitely many others by making use of three procedures \cite{JKSV.Invariants,Vilson.Some}: i) forming arbitrary functions of invariants; 
ii) introducing quotient of derivatives $\mathcal{I}_m \equiv \mathcal{I}_{k}^{\prime}/\mathcal{I}_{l}^{\prime} \equiv \mathrm{d} \mathcal{I}_{k}/\mathrm{d} \mathcal{I}_{l}$; iii) integrating in the sense of indefinite integral $\mathcal{I}_k \equiv \int \mathcal{I}_m\mathcal{I}_{l}^{\prime} \,\mathrm{d}\Phi$. As an example of the first rule, let us define
\begin{equation}
\label{I.4}
\mathcal{I}_{4} \equiv \frac{ \mathcal{I}_{2} }{ \mathcal{I}_{1}^{2} } = \mathrm{e}^{-4\alpha} \mathcal{V} \,.
\end{equation}
Via the second rule, one can make an invariant statement whether a given invariant is dynamical or not. For example, if
\begin{equation}
\label{I.5}
\mathcal{I}_5 \equiv \left( \frac{1}{2} \frac{\mathrm{d} \ln \mathcal{I}_1}{\mathrm{d}\mathcal{I}_3} \right)^2 = \frac{ \left(2\mathcal{A}\alpha^{\,\prime} - \mathcal{A}^\prime \right)^2 }{ 2\mathcal{A}\mathcal{B} + 3 \left( \mathcal{A}^\prime \right)^2 }
\end{equation}
is identically equal to zero, then we have a minimal coupling. Hence $\mathcal{I}_5$ measures the ``strength'' of a non-minimal coupling in an invariant manner. Note also that in the case of the vanishing potential $\mathcal{V}$ of the scalar field this quantity is central in the parametrized post-Newtonian approximation \cite{JKSV.Invariants} and, from field theoretical viewpoint, a (field-dependent) coupling ``constant'' of a scalar interaction mediating between two material bodies \cite{Damour.Nordtvedt}.


\subsubsection{Tensorial invariants}\label{tensorial.invariants}

In addition to scalar invariants we can use the particular transformation properties (\ref{Flanagan.A})-(\ref{Flanagan.alpha}) of the arbitrary functions $\lbrace \mathcal{A}$, $\mathcal{B}$, $\mathcal{V}$, $\alpha \rbrace$ to define an invariant metric \cite{Postma:2014vaa,JKSV.Invariants,Wetterich:2015ccd,Vilson.Some}
\begin{equation}
\label{Invariant.metric1}
\hat{g}_{\mu\nu} \equiv \mathcal{A}(\Phi)g_{\mu\nu} \,.
\end{equation}
Due to the fact that a local rescaling of the metric (\ref{conformal.transformation}) and a redefinition of the scalar field (\ref{field.redefinition}) in principle have nothing to do with coordinate transformations, we conclude that derivatives of the metric $\hat{g}_{\mu\nu}$ w.r.t.\ a spacetime point are invariants in the previous sense as well. Hence, based on the tensor (\ref{Invariant.metric1}), we can construct an arbitrary geometrical object, such that it is manifestly invariant under the transformations (\ref{conformal.transformation})-(\ref{field.redefinition}). However, let us point out that the choice of the invariant metric is not unique. Namely, multiplying the tensor (\ref{Invariant.metric1}) by a scalar invariant does not change its transformation properties. For example \cite{Flanagan},
\begin{equation}
\label{Invariant.metric2}
\bar{g}_{\mu\nu} \equiv \mathrm{e}^{2\alpha(\Phi)}g_{\mu\nu} = \mathcal{I}_{1} \, \hat{g}_{\mu\nu} \,
\end{equation}
also remains invariant under the transformations (\ref{conformal.transformation})-(\ref{field.redefinition}).

Note that the matter fields $\chi$ couple to the latter invariant metric $\bar{g}_{\mu\nu}$ and hence one could consider this to be the invariant Jordan frame metric, which in the Jordan frame $(\alpha \equiv 0)$ functionally coincides with the ordinary metric. Analogously, if one calculates the Ricci tensor for the former invariant metric (\ref{Invariant.metric1}), then in the action (\ref{fl.moju}) the coupling $\mathcal{A}(\Phi)$ could be absorbed into the definition of $\hat{R}\left[ \hat{g}_{\mu\nu} \right]$. Therefore, it is natural to refer to $\hat{g}_{\mu\nu}$ as the invariant Einstein frame metric, which in the Einstein frame $(\mathcal{A} \equiv 1)$ coincides with the ordinary metric. In what follows, we shall drop the prefix invariant, because without foreknowledge about the transformation properties one cannot determine whether a given metric in these frames is invariant under (\ref{conformal.transformation})-(\ref{field.redefinition}) or not. Indeed, even the relation (\ref{Invariant.metric2}) between the two invariant metrics can be seen as an ordinary conformal transformation between the Jordan and Einstein frame \cite{Vilson.Some}. Nevertheless, we shall use the invariant formulation, which explicitly contains all four arbitrary functions $\lbrace \mathcal{A}$, $\mathcal{B}$, $\mathcal{V}$, $\alpha \rbrace$, in order to facilitate easy comparison of different parametrizations.


\subsection{Equations of motion on a FLRW background}


\subsubsection{Invariants in FLRW cosmology}\label{Invariants.in.FLRW.cosmology}

In the current paper we shall consider the equations of motion on a flat FLRW background, where the line element is given by
\begin{equation}
\label{FLRW.line.element}
\mathrm{d}s^2 \equiv g_{\mu\nu}\mathrm{d}x^\mu \mathrm{d}x^\nu = -\mathrm{d}t^2 + \left(a(t)\right)^2 \delta_{ij}\mathrm{d}x^i \mathrm{d}x^j \,.
\end{equation}
Due to homogeneity and isotropy assumptions underlying the FLRW cosmology, the scalar field and the invariants introduced in subsection~\ref{Subsec:Invariant.quantities} may only depend on the cosmological time, i.e.\ $\mathcal{I}_{i} = \mathcal{I}_{i}(t)$ etc.

It is straightforward to put the invariant metric $\hat{g}_{\mu\nu}$ (\ref{Invariant.metric1}) also in a FLRW form by performing the following coordinate transformation and scale factor redefinition
\begin{equation}
\label{coordinate.transformation.t.to.t.hat}
\frac{\mathrm{d} \phantom{\hat{t}}}{\mathrm{d}\hat{t}} \equiv \frac{1}{\sqrt{\mathcal{A}}}\frac{\mathrm{d} \phantom{{t}}}{\mathrm{d}t} \,, \qquad
\hat{a}(\hat{t} ) \equiv \sqrt{\mathcal{A}} \, a(t) \,.
\end{equation}
Note that $\hat{a}^2$ is in principle the so-called Lorenz-Petzold variable \cite{LorenzPetzold}. An analogous transformation, namely
\begin{equation}
\label{coordinate.transformation.t.to.t.bar}
\frac{\mathrm{d} \phantom{\hat{t}}}{\mathrm{d}\bar{t}} \equiv \frac{1}{\mathrm{e}^{\alpha}}\frac{\mathrm{d} \phantom{{t}}}{\mathrm{d}t}= \frac{1}{\sqrt{\mathcal{I}_1}}\frac{\mathrm{d} \phantom{{t}}}{\mathrm{d}\hat{t}} \,, \qquad
\bar{a}(\bar{t} ) \equiv \mathrm{e}^{\alpha} \, a(t) =\sqrt{\mathcal{I}_1}\hat{a}(\hat{t} )\,
\end{equation}
takes the invariant metric (\ref{Invariant.metric2}) to the FLRW form as well, but now with a different time parameter. 

The invariants preserve their numerical value under the conformal transformation (\ref{conformal.transformation}) and scalar field reparametrization (\ref{field.redefinition}) for particular values of the coordinates, if these do not change. However, for FLRW cosmology, it is well known that in order to avoid introducing a lapse function, each conformal transformation is followed by a time rescaling. The latter causes the invariants to transform as scalar functions and hence their numerical value for a particular value of the cosmological time is not preserved, i.e.\ 
\begin{equation}
\left. \mathcal{I}_{i}(t) \right|_{ t = t_0 } \neq \left. \mathcal{I}_{i}(\hat{t}) \right|_{ \hat{t} = t_0 } \neq \left. \mathcal{I}_{i}(\bar{t}) \right|_{ \bar{t} = t_0 } \,.
\end{equation}
In order to avoid confusion, one may consider the invariants to be functions of the conformal time $\mathrm{d}\tau \equiv \mathrm{d}t / a $, because it is manifestly invariant.


\subsubsection{Einstein frame equations}

Varying the action (\ref{fl.moju}) with respect to the metric and the scalar field yields equations of motion which can be specified for the FLRW background (\ref{FLRW.line.element}).

The Hubble parameter $\hat{H}$ calculated in terms of the invariant metric $\hat{g}_{\mu\nu}$ (\ref{Invariant.metric1}) and the Hubble parameter $H$ calculated in the frame defined by $g_{\mu\nu}$ (\ref{FLRW.line.element}) are related by 
\begin{equation}
\label{hubble.relation}
\hat{H} \equiv \frac{1}{\hat{a}} \frac{ \mathrm{d} \hat{a} }{ \mathrm{d}\hat{t} }= \frac{1}{\sqrt{\mathcal{A}}}\left(H + \frac{1}{2}\frac{\mathcal{A}^\prime}{\mathcal{A}} \frac{ \mathrm{d}\Phi }{ \mathrm{d} t}\right) \,.
\end{equation}
Field equations in terms of invariants (\ref{I.1.I.2})-(\ref{I.5}) and cosmological time coordinate $\hat{t}$ (\ref{coordinate.transformation.t.to.t.hat}) read
\begin{subequations}
%
\begin{eqnarray}
\label{Friedmanns.constraint.in.terms.of.invariants} 
\hat{H}^2 &=& \frac{1}{3}\left( \frac{ \mathrm{d}\mathcal{I}_3 }{ \mathrm{d}\hat{t} } \right)^2 + \frac{1}{3\ell^2}\mathcal{I}_2 + \frac{\kappa^2}{3} \hat{\rho} \,, \\
\label{Friedmanns.second.equation.in.terms.of.invariants}
\frac{ \mathrm{d}\hat{H}  }{ \mathrm{d}\hat{t} }   &=&  -\left(\frac{ \mathrm{d} \mathcal{I}_3 }{\mathrm{d}\hat{t} } \right)^2 -\frac{\kappa^2}{2}\left( \hat{\rho}+\hat{p}\right)   \,, \\ 
\label{Friedmanns.scalar.field.equation} 
\frac{ \mathrm{d}^2 \mathcal{I}_3 }{ \mathrm{d}\hat{t}^2 }    &=& -3\hat{H} \frac{ \mathrm{d} \mathcal{I}_3 }{ \mathrm{d}\hat{t} }  - \frac{1}{2\ell^2}\frac{ \mathrm{d}\mathcal{I}_2 }{ \mathrm{d}\mathcal{I}_3 }
+\frac{\kappa^2}{4} \frac{ \mathrm{d}\ln\mathcal{I}_1 }{ \mathrm{d} \mathcal{I}_3}(-\hat{\rho}+3\hat{p}) \,.
\end{eqnarray}
%
\end{subequations}
Here $\hat{\rho} \equiv \rho \, \mathcal{A}^{-2}$ and $\hat{p} \equiv p \, \mathcal{A}^{-2}$ represent the invariant energy density and pressure for the matter content, respectively. For more details on the form of field equations in the general parametrization and their transformation properties consider \cite{JKSV.Invariants,Jarv:2015kga}. 


\subsubsection{Jordan frame equations}

Let us rewrite the system (\ref{Friedmanns.constraint.in.terms.of.invariants})-(\ref{Friedmanns.scalar.field.equation}) using the invariant metric $\bar{g}_{\mu\nu}=\mathrm{e}^{2 \alpha}g_{\mu\nu}$ (\ref{Invariant.metric2}). 
Note that $\bar{g}_{\mu\nu}$ is the metric which couples to matter in the action functional (\ref{fl.moju}). If the frame is specified as the Jordan one, 
$\alpha = 0$, 
then it coincides with the ordinary Jordan frame and hence trajectories of free particles are geodesics. If specified as the Einstein frame, $\mathcal{A} = 1, \alpha \not= 0$, equations of motion of free particles obtain a force term on r.h.s.\ determined by the scalar field. Hence the invariant Jordan frame incorporates Einstein and Jordan frames in their familiar forms. Chiba and Yamaguchi
\cite{Chiba:2013mha} studied in detail relations between cosmological observables in Jordan and Einstein  frames and presented a Jordan-Einstein dictionary. All this information is encoded also in our invariant Jordan frame formalism.  

Some authors refer to the Jordan frame as a frame where the Ricci scalar is multiplied by a dynamical function of the scalar field (dynamical $\mathcal{A}$), and do not explicitly require the minimal coupling to matter fields, e.g.\ \cite{Burns.et.al}. Then one can speak of making conformal transformations between Jordan frames, whereas by our definition, the Jordan frame is uniquely fixed as the frame where matter is minimally coupled. In the latter case there exist still different parametrizations for the Jordan frame since one can redefine the scalar field.

Similarly to (\ref{coordinate.transformation.t.to.t.hat}) we can put the invariant metric $\bar{g}_{\mu\nu}$ in a FLRW form by performing the coordinate transformation and the scale factor redefinition as in (\ref{coordinate.transformation.t.to.t.bar}). We can define the Hubble parameter $\bar{H}$ that corresponds to the invariant metric $\bar{g}_{\mu\nu}$ by
\begin{equation}
	\label{hubble.relation.JF}
	\bar{H} \equiv \frac{1}{\bar{a}} \frac{\mathrm{d} \bar{a} }{\mathrm{d}\bar{t}}= \mathrm{e}^{-\alpha} \left(H + \alpha^{\,\prime} \frac{ \mathrm{d}\Phi }{ \mathrm{d} t } \right)= 
	\frac{1}{\sqrt{\mathcal{I}_1}}\left(\hat{H} + \frac{1}{2}\frac{ \mathrm{d} \ln \mathcal{I}_1}{ \mathrm{d} \hat{t} }\right)
	\,.
\end{equation}

We have introduced three different Hubble parameters, $H$, $\hat{H}$, $\bar{H}$. It is 
often claimed that the observable Hubble parameter belongs to the ordinary Jordan frame which coincides with the invariant Jordan frame upon suitable specification of the two functional degrees of freedom. Accordingly, the relevant information about observables is contained in $\bar{H}$. In \cite{Chiba:2013mha} it was demonstrated that the measurable Hubble parameter in the Einstein frame $H_{E}$ should simply be defined by 
$H_{E}=\sqrt{\mathcal{I}_1}\bar{H}$. Comparing with (\ref{hubble.relation.JF}) we see that $H_{E}$ is not equal to $\hat{H}$, that is a quantity defined by the FLRW metric in the Einstein frame, and the latter one is declared to be non-measurable.

Using (\ref{coordinate.transformation.t.to.t.bar}) and (\ref{hubble.relation.JF}) it is straightforward to recast the system (\ref{Friedmanns.constraint.in.terms.of.invariants})-(\ref{Friedmanns.scalar.field.equation}) in the following form:
\begin{subequations}
%
\begin{eqnarray}
	\label{Friedmann1.JF} 
	\bar{H}^2 &=& \frac{1}{3}\left( \frac{ \mathrm{d} \mathcal{I}_3 }{ \mathrm{d}\bar{t} } \right)^2 
	+\bar{H}\frac{ \mathrm{d} \ln \mathcal{I}_1 }{ \mathrm{d} \bar{t}}-\frac{1}{4}\left(\frac{ \mathrm{d} \ln \mathcal{I}_1 }{ \mathrm{d} \bar{t}} \right)^2
	+ \frac{1}{3\ell^2}\frac{ \mathcal{I}_2 }{\mathcal{I}_1} +\frac{\kappa^2}{3}  \mathcal{I}_1\bar{\rho} \,, \\
	\label{Friedmann2.JF}
	\frac{ \mathrm{d} \bar{H} }{ \mathrm{d}\bar{t} }    &=& - \frac{1}{2} \bar{H}\frac{ \mathrm{d} \ln \mathcal{I}_1 }{ \mathrm{d} \bar{t}} + \frac{1}{4} \left(\frac{ \mathrm{d} \ln \mathcal{I}_1 }{ \mathrm{d} \bar{t}} \right)^2 -\left(\frac{\mathrm{d} \mathcal{I}_3 }{ \mathrm{d}\bar{t} } \right)^2 +\frac{1}{2} \frac{ \mathrm{d}^2 \ln \mathcal{I}_1 }{ \mathrm{d} \bar{t}^2} -\frac{\kappa^2}{2} \mathcal{I}_1 \left( \bar{\rho} + \bar{p} \right)   \,, \\
	\label{Friedmann3.JF} 
	 \frac{ \mathrm{d}^2 \mathcal{I}_3 }{ \mathrm{d}\bar{t}^2 }  &=& \left(-3\bar{H} + \frac{ \mathrm{d} \ln \mathcal{I}_1 }{ \mathrm{d} \bar{t}} \right) \frac{ \mathrm{d} \mathcal{I}_3 }{ \mathrm{d} \bar{t} } - \frac{1}{2\ell^2 \mathcal{I}_1} \frac{ \mathrm{d}\mathcal{I}_2 }{ \mathrm{d}\mathcal{I}_3 } + \frac{\kappa^2}{4} \mathcal{I}_1 \frac{ \mathrm{d} \ln \mathcal{I}_1 }{ \mathrm{d} \mathcal{I}_3 }( -\bar{\rho} + 3\bar{p} ) 
	\,.
\end{eqnarray}
%
\end{subequations}
Here $\bar{\rho} \equiv \rho \, \mathrm{e}^{-4\alpha} = \hat{\rho} \, \mathcal{I}_{1}^{\, -2}$ and $\bar{p} \equiv p \, \mathrm{e}^{-4\alpha} = \hat{p} \, \mathcal{I}_{1}^{\, -2}$ denote the invariant energy density and pressure for the matter content that correspond to those quantities in the Jordan frame, respectively.
The invariant $\bar{H}$ (\ref{hubble.relation.JF}) represents the Hubble parameter and $\bar{t}$ (\ref{coordinate.transformation.t.to.t.bar}) denotes the time coordinate in the Jordan frame for the FLRW background. Thus one can claim that (\ref{Friedmann1.JF})-(\ref{Friedmann3.JF}) are in the same form as the field equations in the Jordan frame. Note that both $\mathcal{I}_{3}$ and $\mathcal{I}_1$ are differentiated w.r.t.\ time coordinate and hence the scalar field is unspecified. This does not concern us, because we are interested in the behaviour of the Hubble parameter $\bar{H}$. Details about invariants in different parametrizations are presented in table \ref{Table} in appendix \ref{App:Invariants.in.different.parametrizations}.  


\section{Slow-roll }\label{Sec:Slow.roll}


\subsection{Slow-roll in the minimally coupled theory \texorpdfstring{\cite{Lyth:2009zz}}{}}

Let us consider the theory given by the action (\ref{nmcsf}) and assume the FLRW metric (\ref{FLRW.line.element}). A common requirement for inflation is that the potential dominates over the kinetic term $\mathcal{V}(\phi)~\gg~\dot{\phi}^2$. The qualitative description is realized through the dimensionless Hubble slow-roll (HSR) parameters usually defined as
\begin{equation}
	\label{hsr}
	\epsilon_{_H} \equiv -\frac{\dot{H}}{H^2}\, , \qquad \qquad
	\eta_{_H} \equiv - \frac{1}{H} \, \frac{\ddot{\phi}}{\dot{\phi}} \,.
\end{equation}
The condition $\epsilon_{_H} \ll 1$ implies that expansion is nearly de Sitter like, and $\eta_{_H} \ll 1$ guarantees sufficient duration of inflation.
One can define also the potential slow-roll (PSR) parameters $\epsilon_{_V}$ and $\eta_{_V}$ as follows
\begin{subequations}
%
\begin{eqnarray}
\label{PSRnm1}
\epsilon_{_V} &\equiv& \frac{1}{4} \left(\frac{\mathcal{V'}}{\mathcal{V}} \right)^2 = \epsilon_{_H} \left( \frac{ 3 - \eta_{_H} }{ 3 - \epsilon_{_H} } \right)^2 \approx \epsilon_{_H}  \,, \\
\label{PSRnm2}
\eta_{_{V}} &\equiv&  \frac{1}{2} \left(\frac{\mathcal{V''}}{\mathcal{V}} \right) \approx
\epsilon_{_H} + \eta_{_H} \,.
\end{eqnarray}
%
\end{subequations}
The equality in (\ref{PSRnm1}) follows directly from equations of motion. For (\ref{PSRnm2}) further manipulation of the equations is used and the exact correspondence is given in \cite{Liddle:1994dx}. In the approximated equalities, only linear terms in HSR parameters are kept and the result is of the first order.

In order to characterize the evolution of primordial perturbations, one may introduce the scalar tilt $n_{_\mathrm{S}}$, tensor tilt $n_{_\mathrm{T}}$
and tensor-to-scalar ratio $r$ \cite{Lyth:2009zz}, which can be expressed to first order in terms of the slow-roll parameters (\ref{hsr})
\begin{equation}
\label{observables.in.minimally.coupled.theory}
n_{_\mathrm{S}} = 1 - 4 \epsilon_{_H} + 2 \eta_{_H} 
\,, \qquad \qquad
n_{_\mathrm{T}} = - 2 \epsilon_{_H} 
\,, \qquad \qquad
r = 16 \epsilon_{_H} 
\,.
\end{equation}


\subsection{Slow-roll in the Einstein frame}

In order to study slow-roll in a general scalar-tensor theory, let us first consider the system (\ref{Friedmanns.constraint.in.terms.of.invariants})-(\ref{Friedmanns.scalar.field.equation}) without matter terms
\begin{subequations}
%
\begin{eqnarray}
\label{Friedmann1.EF} 
\hat{H}^2 &=& \frac{1}{3} \left( \frac{ \mathrm{d} \mathcal{I}_3  }{ \mathrm{d}\hat{t} } \right)^2 + \frac{1}{3\ell^2} \mathcal{I}_2  \,, \\
\label{Friedmann2.EF}
\frac{ \mathrm{d} \hat{H} }{ \mathrm{d}\hat{t} } &=&  -\left( \frac{ \mathrm{d} \mathcal{I}_3 }{ \mathrm{d}\hat{t} } \right)^2  \,, \\
\label{Friedmann3.EF} 
\frac{ \mathrm{d}^2 \mathcal{I}_3 }{ \mathrm{d}\hat{t}^2 }  &=& -3\hat{H} \frac{\mathrm{d} \mathcal{I}_3 }{ \mathrm{d}\hat{t} } - \frac{1}{2\ell^2} \frac{\mathrm{d}\mathcal{I}_2}{ \mathrm{d}\mathcal{I}_3 }
\,.
\end{eqnarray}
%
\end{subequations}
This system, written down in terms of invariants, is formally the same as the one in the canonical Einstein frame with $\mathcal{I}_2$ in the role of the potential and $\mathcal{I}_3$ in the role of the scalar field, and if we choose to work within the ordinary Einstein frame, then all terms reduce to the corresponding ones without any mixing \cite{Vilson.Some}. Note that from the viewpoint of transformation properties, neglecting matter is a covariant procedure w.r.t.\ the conformal transformation (\ref{conformal.transformation}) and scalar field redefinition (\ref{field.redefinition}), because the energy-momentum tensor itself only gains a finite multiplier under these transformations. Due to neglecting matter, the equations in the Einstein frame are in the same form as the equations in general relativity with a minimally coupled scalar field. Based on this well-known fact, we start by generalizing the results obtained in the minimal case to the non-minimal case in a straightforward manner, i.e.\ we impose the standard slow-roll conditions for the invariant quantities in (\ref{Friedmann1.EF})-(\ref{Friedmann3.EF}) \cite{Liddle:1994dx}.

Let us define the following invariant parameters for the HSR \cite{Liddle:1994dx} 
\begin{equation}
\label{epsilon0.EF}
\hat{\epsilon}_{0} \equiv - \frac{1}{\hat{H}^2} \frac{ \mathrm{d}\hat{H} }{ \mathrm{d}\hat{t} } = - \frac{ \mathrm{d} \ln \hat{H} }{ \mathrm{d} \ln \hat{a} } 
\,, \qquad \qquad
\hat{\eta} \equiv -\left( \hat{H} \frac{ \mathrm{d} \mathcal{I}_{3} }{ \mathrm{d} \hat{t} } \right)^{-1} \frac{ \mathrm{d}^2 \mathcal{I}_{3} }{ \mathrm{d} \hat{t}^2 } \,.
\end{equation}
The condition for accelerated expansion in the Einstein frame reads precisely $\hat{\epsilon}_{0} < 1$. While considering slow-roll inflation we shall assume $\hat{\epsilon}_{0} \ll 1$, which states that the expansion rate in the Einstein frame is almost exponential. By making use of recursive prescription analogously to \cite{vandeBruck:2015gjd,Torres:1996fr}, let us additionally define invariant parameters
\begin{subequations}
%
\begin{eqnarray}
\label{kappa.EF}
\hat{\kappa}_{0} &\equiv& \frac{1}{ \hat{H}^2 } \left( \frac{ \mathrm{d}\mathcal{I}_3 }{ \mathrm{d}\hat{t} }\right)^2 = \left( \frac{ \mathrm{d} \mathcal{I}_{3} }{ \mathrm{d} \ln \hat{a} } \right)^2 \,, \qquad
\hat{\kappa}_{1} \equiv \frac{1}{ \hat{H} \hat{\kappa}_{0} } \frac{ \mathrm{d}\hat{\kappa}_{0} }{ \mathrm{d}\hat{t} } = \frac{ \mathrm{d} \ln \hat{\kappa}_{0} }{ \mathrm{d} \ln \hat{a} } = 2 \left( - \hat{\eta} + \hat{\epsilon}_{0} \right)
\,, \\
\label{kappa.i.EF}
\hat{\kappa}_{i + 1} &\equiv& \frac{ 1 }{ \hat{H} \hat{\kappa}_{i} } \frac{ \mathrm{d} \hat{\kappa}_{i} }{ \mathrm{d} \hat{t} } = \frac{ \mathrm{d} \ln \hat{\kappa}_{i} }{ \mathrm{d} \ln \hat{a} } \,.
\end{eqnarray}
%
\end{subequations}
Using the definitions (\ref{epsilon0.EF}) and (\ref{kappa.EF}) we can write the system (\ref{Friedmann1.EF})-(\ref{Friedmann3.EF}) in the following form
\vspace{-0.6cm}
\begin{subequations}
%
\begin{eqnarray}
\label{Friedmann1.EF1}
\mathcal{I}_2 \ell^{-2} &=&  \hat{H}^2 \left(3 - \hat{\kappa}_{0} \right)\,, \\
\label{Friedmann2.EF1}
\hat{\kappa}_{0} &=&  \hat{\epsilon}_{0}   \,, \\
\label{Friedmann3.EF1} 
 \hat{H} \frac{ \mathrm{d} \mathcal{I}_{3} }{ \mathrm{d} \hat{t} } \left( 3 - \hat{\epsilon}_{0} + \frac{1}{2} \hat{\kappa}_{1} \right) &=& - \frac{1}{2\ell^2} \frac{ \mathrm{d} \mathcal{I}_2 }{ \mathrm{d} \mathcal{I}_3 } 
 \,.
\end{eqnarray}
%
\end{subequations}
Slow-roll inflation corresponds to
\begin{equation}
\label{slow.roll.EF}
| \hat{\kappa}_{0} |\ll 1 \,, \qquad \qquad 
|\hat{\kappa}_{1} |\ll 1 \,,
\end{equation}
and considering this regime the system (\ref{Friedmann1.EF1})-(\ref{Friedmann3.EF1}) can be approximated as
\begin{equation}\label{slow.roll.EF.approximation}
\mathcal{I}_2 \ell^{-2}\approx 3\hat{H}^2 \,, \qquad \qquad
3\hat{H} \frac{ \mathrm{d} \mathcal{I}_{3} }{ \mathrm{d} \hat{t} } \approx - \frac{1}{2\ell^2} \frac{ \mathrm{d} \mathcal{I}_2 }{ \mathrm{d} \mathcal{I}_3 }
\,.
\end{equation}
Due to the equality in (\ref{Friedmann2.EF1}) the slow-roll conditions (\ref{slow.roll.EF}) are essentially defined in terms of the derivatives of the Hubble parameter. Via straightforward manipulations we can use the approximated system (\ref{slow.roll.EF.approximation}) in order to rephrase the conditions (\ref{slow.roll.EF}) in terms of the derivatives of the invariant potential $\mathcal{I}_2$ and obtain relations between PSR and HSR parameters up to first order in slow-roll parameters
\begin{equation}
\label{parameters.in.terms.of.potential}
\hat{\kappa}^{(\mathcal{V})}_{0} \equiv \frac{1}{ 4 \mathcal{I}_{2}^{2} } \left( \frac{ \mathrm{d} \mathcal{I}_2 }{\mathrm{d} \mathcal{I}_3} \right)^2 \approx \hat{\kappa}_{0} \,, \qquad \qquad
\hat{\kappa}^{(\mathcal{V})}_{1} \equiv 4 \hat{\kappa}^{(\mathcal{V})}_{0} - \frac{1}{ \mathcal{I}_2 } \frac{ \mathrm{d}^2 \mathcal{I}_{2} }{ \mathrm{d} \mathcal{I}_{3}^{2} } \approx \hat{\kappa}_{1} \,.
\end{equation}
For second order correspondence see (\ref{HSR.vr.PSR.so1})-(\ref{HSR.vr.PSR.so2}). One might conclude that the slow-roll regime is formally defined in exactly the same way as in the case of the standard minimal slow-roll. However, note that the (spacetime independent) units are defined w.r.t.\ geometry $\hat{g}_{\mu\nu}$ (cf.\ with the final part of subsection~2.2 in \cite{Flanagan}). 

Let us suppose that one wants to use the units defined via the matter fields instead. Then, in the spirit of Dicke \cite{Dicke:1962gz}, one should also take into account that for a non-minimally coupled theory in the Einstein frame these units are spacetime point dependent, despite the fact that we are using the invariant metric and dropped the matter fields. This is due to the fact that the metric $\bar{g}_{\mu\nu}$ (\ref{Invariant.metric2}), to which the matter fields should couple, is also invariant and related to the geometrical one $\hat{g}_{\mu\nu}$ via a conformal transformation (\ref{Invariant.metric2}). Hence, in the case of the ``matter'' units, slow-roll in the Einstein frame could be induced by a suitable time dependence of these units. In order to avoid this one should introduce extra conditions. Let us define
\begin{subequations}
%
\begin{eqnarray}
\label{lambda.0.EF}
\hat{\lambda}_{0} &\equiv& \frac{1}{2\hat{H}} \frac{ \mathrm{d} \ln \mathcal{I}_1}{ \mathrm{d} \hat{t}} = \frac{1}{2} \frac{ \mathrm{d} \ln \mathcal{I}_{1} }{ \mathrm{d} \ln \hat{a} } = \pm \sqrt{\mathcal{I}_5} \sqrt{ \hat{\kappa}_0 }\,, \\
\label{lambda.1.EF}
\hat{\lambda}_{1} &\equiv& \frac{1}{\hat{H} \hat{\lambda}_{0}} \frac{ \mathrm{d} \hat{\lambda}_{0} }{ \mathrm{d} \hat{t} } = \frac{ \mathrm{d} \ln \hat{\lambda}_{0} }{ \mathrm{d} \ln \hat{a} } = \pm \frac{1}{2} \frac{ \mathrm{d} \ln \mathcal{I}_{5} }{ \mathrm{d} \mathcal{I}_{3} } \sqrt{ \hat{\kappa}_{0} } + \frac{1}{2} \hat{\kappa}_{1}
\,, \\
\label{lambda.i.EF}
\hat{\lambda}_{i + 1} &\equiv& \frac{1}{ \hat{H} \hat{\lambda}_{i} } \frac{ \mathrm{d} \hat{\lambda}_{i} }{ \mathrm{d} \hat{t} } = \frac{ \mathrm{d} \ln \hat{\lambda}_{i} }{ \mathrm{d} \ln \hat{a} } \,,
\end{eqnarray}
%
\end{subequations}
where we used the definition (\ref{I.5}) for $\mathcal{I}_5$. Imposing the conditions
\begin{equation}
\label{slow.roll.EF1}
| \hat{\lambda}_{0} | \ll 1 \,, \qquad \qquad
| \hat{\lambda}_{1} |  \ll 1 \,,
\end{equation}
leads to slow-roll w.r.t.\ the ``matter'' units as well. In the case of a minimally coupled scalar field ($\mathcal{I}_5 \equiv 0$) these extra conditions are manifestly fulfilled as they should.

Notice that we defined slow-roll in the Einstein frame by imposing conditions (\ref{slow.roll.EF}). In order to have this regime sustained for sufficiently long one should have $| \hat{\kappa}_{i} | \ll 1$ as well for the parameters defined in  (\ref{kappa.i.EF}). These parameters appear if we take higher time derivatives of the system (\ref{Friedmann1.EF1})-(\ref{Friedmann3.EF1}) and they form a so-called slow-roll hierarchy \cite{Liddle:1994dx}. They are needed for obtaining higher-order results in slow-roll parameters. Similar situation applies for slow-roll w.r.t.\ the ``matter'' units, where in addition to (\ref{slow.roll.EF1}) one imposes $| \hat{\lambda}_{i} | \ll 1$ for the parameters defined in  (\ref{lambda.i.EF}).

In the literature occasionally some invariant metric, often constructed on dimensional grounds, is used to make invariant claims \cite{Postma:2014vaa,Wetterich:2015ccd}. However, let us point out that there is an ambiguity in the choice of the invariant metric which spoils the possibility to solve problems easily.


\subsection{Slow-roll in the Jordan frame}

Similarly to the Einstein frame we can define the invariant parameters
\begin{equation}
\label{epsilon0.JF}
\bar{\epsilon}_{0} \equiv -\frac{1}{\bar{H}^2} \frac{ \mathrm{d}\bar{H} }{ \mathrm{d}\bar{t} } = - \frac{ \mathrm{d} \ln \bar{H} }{ \mathrm{d} \ln \bar{a} }  \,, \qquad \qquad 
\bar{\eta} \equiv -\left( \bar{H} \frac{ \mathrm{d} \mathcal{I}_{3} }{ \mathrm{d} \bar{t} } \right)^{-1} \frac{ \mathrm{d}^2 \mathcal{I}_{3} }{ \mathrm{d} \bar{t}^2 } 
\,.
\end{equation}
Now the condition for inflation in the Jordan frame reads $\bar{\epsilon}_{0} < 1$ and to obtain almost exponential expansion rate one restricts $\bar{\epsilon}_{0} \ll 1$. For slow-roll in the Jordan frame we define similarly to (\ref{kappa.EF})-(\ref{kappa.i.EF}) the following invariant parameters
\begin{subequations}
%
\begin{eqnarray}
\label{kappa.JF}
\bar{\kappa}_{0} &\equiv& \frac{1}{\bar{H}^2} \left( \frac{ \mathrm{d} \mathcal{I}_3 }{ \mathrm{d} \bar{t} } \right)^2 = \left( \frac{ \mathrm{d} \mathcal{I}_{3} }{ \mathrm{d} \ln \bar{a} } \right)^2 \, , \qquad
\bar{\kappa}_{1} \equiv \frac{1}{ \bar{H} \bar{\kappa}_{0} } \frac{ \mathrm{d} \bar{\kappa}_{0} }{ \mathrm{d} \bar{t} } = \frac{ \mathrm{d} \ln \bar{\kappa}_{0} }{ \mathrm{d} \ln \bar{a} } = 2 \left( - \bar{\eta} + \bar{\epsilon}_{0} \right)
\,, \\
\label{kappa.i.JF}
\bar{\kappa}_{i+1} &\equiv& \frac{1}{ \bar{H} \bar{\kappa}_{i} } \frac{ \mathrm{d} \bar{\kappa}_{i} }{ \mathrm{d} \bar{t} } = \frac{ \mathrm{d} \ln \bar{\kappa}_{i} }{ \mathrm{d} \ln \bar{a} } \,.
\end{eqnarray}
%
\end{subequations}
Let us additionally define invariant parameters
\begin{subequations}
%
\begin{eqnarray}
\label{lambda.0.JF}
\bar{\lambda}_{0} &\equiv& \frac{1}{2\bar{H}} \frac{ \mathrm{d} \ln \mathcal{I}_1 }{ \mathrm{d} \bar{t} } = \frac{1}{2} \frac{ \mathrm{d} \ln \mathcal{I}_{1} }{ \mathrm{d} \ln \bar{a} } = \pm \sqrt{\mathcal{I}_5} \sqrt{\bar{\kappa}_{0} } \, , \\
\label{lambda.1.JF}
\bar{\lambda}_{1} &\equiv& \frac{1}{\bar{H} \bar{\lambda}_{0} } \frac{ \mathrm{d} \bar{\lambda}_{0} }{ \mathrm{d} \bar{t} } = \frac{ \mathrm{d} \ln \bar{\lambda}_{0} }{ \mathrm{d} \ln \bar{a} } =
\pm \frac{1}{2} \frac{ \mathrm{d} \ln \mathcal{I}_{5} }{ \mathrm{d} \mathcal{I}_{3} } \sqrt{ \bar{\kappa}_{0} } + \frac{1}{2} \bar{\kappa}_{1} 
\,, \\
\label{lambda.i.JF}
\bar{\lambda}_{i+1} &\equiv& \frac{ 1 }{ \bar{H} \bar{\lambda}_{i} } \frac{ \mathrm{d} \bar{\lambda}_{i} }{ \mathrm{d} \bar{t} } = \frac{ \mathrm{d} \ln \bar{\lambda}_{i} }{ \mathrm{d} \ln \bar{a} } \,,
\end{eqnarray}
%
\end{subequations}
which, as in the case of the Einstein frame, measure the non-minimal coupling (see also \cite{vandeBruck:2015gjd,Morris:2001ad,Chiba:2008ia,Noh:2001ia}). Using the definitions (\ref{epsilon0.JF})-(\ref{lambda.i.JF}) we can recast the system (\ref{Friedmann1.JF})-(\ref{Friedmann3.JF}) without matter in the following form
\vspace{-0.2cm} 
\begin{subequations}
%
\begin{eqnarray}
\label{Friedmann1.JF1}
\mathcal{I}_2 \ell^{-2} &=&  \bar{H}^2 \mathcal{I}_1  \left(3 - \bar{\kappa}_{0} - 3 \bar{\lambda}_{0} \left( 2 - \bar{\lambda}_{0} \right) \right) \,, \\
\label{Friedmann2.JF1}
\bar{\epsilon}_{0} &=& \bar{\kappa}_{0}  +  \bar{\lambda}_{0} \left( 1 + \bar{\epsilon}_{0} -  \bar{\lambda}_{0} - \bar{\lambda}_{1} \right)  \,, \\
\label{Friedmann3.JF1} 
- \frac{1}{2\ell^2 \mathcal{I}_1 } \frac{ \mathrm{d} \mathcal{I}_2 }{ \mathrm{d} \mathcal{I}_3 }  &=& \bar{H} \frac{ \mathrm{d} \mathcal{I}_{3} }{ \mathrm{d} \bar{t} } \left( 3 - \bar{\epsilon}_{0} +\frac{1}{2} \bar{\kappa}_{1} - 2 \bar{\lambda}_{0} \right)
 \,.
\end{eqnarray}
%
\end{subequations}
By imposing
\begin{equation}
\label{slow.roll.JF}
| \bar{\kappa}_{0} |\ll 1 \,, \qquad
| \bar{\kappa}_{1} |\ll 1
\,, \qquad
| \bar{\lambda}_{0} |\ll 1 \,, \qquad
| \bar{\lambda}_{1} |\ll 1 \,,
\end{equation}
it follows from (\ref{Friedmann2.JF1}) that $| \bar{\epsilon}_{0} | \ll 1$. In this regime the system (\ref{Friedmann1.JF1})-(\ref{Friedmann3.JF1}) can be approximated as
\begin{equation}
\label{slow.roll.JF.approximation}
\mathcal{I}_2 \ell^{-2}\approx 3\bar{H}^2 \mathcal{I}_1 \,, \qquad  \qquad
3 \bar{H} \frac{ \mathrm{d} \mathcal{I}_{3} }{ \mathrm{d} \bar{t} } \approx - \frac{1}{2\ell^2 \mathcal{I}_1 } \frac{ \mathrm{d} \mathcal{I}_2 }{ \mathrm{d} \mathcal{I}_3 }
\,.
\end{equation}
Let us point out that via such an approach, if we do not distinguish between $\hat{H}$ and $\bar{H}$, the approximate equations (\ref{slow.roll.JF.approximation}) of the system (\ref{Friedmann1.JF1})-(\ref{Friedmann3.JF1}) are almost the same as the equations (\ref{slow.roll.EF.approximation}), which approximate the system (\ref{Friedmann1.EF1})-(\ref{Friedmann3.EF1}) in the Einstein frame, while the distinguishing multiplier $\mathcal{I}_{1}$ is due to the change of units.

Analogously to the Einstein frame correspondence (\ref{parameters.in.terms.of.potential}), one can use the approximated system (\ref{slow.roll.JF.approximation}) in order to write the slow-roll parameters up to first order via the invariant potential $\mathcal{I}_2$ and its derivatives as
\begin{equation}
\bar{\kappa}_{0}^{(\mathcal{V})} \equiv \frac{1}{4 \mathcal{I}_{2}^{2}} \left( \frac{ \mathrm{d} \mathcal{I}_2 }{ \mathrm{d} \mathcal{I}_3 } \right)^2 \approx \bar{\kappa}_{0} \,, \qquad \qquad
\bar{\kappa}_{1}^{(\mathcal{V})} \equiv 4 \bar{\kappa}_{0}^{(\mathcal{V})} - \frac{1}{ \mathcal{I}_2 } \frac{ \mathrm{d}^2 \mathcal{I}_{2} }{ \mathrm{d} \mathcal{I}_{3}^{2} } \approx \bar{\kappa}_{1} \,.
\end{equation}
Note that the correspondence between HSR and PSR parameters is formally the same as in the Einstein frame (\ref{parameters.in.terms.of.potential}), and that the definitions of the PSR parameters are identical in both frames. In the Jordan frame the parameters $\bar{\kappa}_{i}$ are not identical to $\bar{\epsilon}_{i}$, due to additional terms in the acceleration equation (\ref{Friedmann2.JF1}) that are not present in the corresponding equation (\ref{Friedmann2.EF1}), written down in the Einstein frame ($\hat{\epsilon}_{i}=\hat{\kappa}_{i}$). These extra terms vanish, if the scalar field is minimally coupled, i.e.\ $\mathcal{I}_{5} \equiv 0$, and for the latter case indeed in the Jordan frame we also have pure Hubble slow-roll, which is consistent with the fact that for a minimally coupled scalar field these two frames coincide.


\subsection{Relation between the slow-roll parameters in different frames}

Let us compare the formulations of the slow-roll regimes in the Einstein and Jordan frames. First, note that imposing (\ref{slow.roll.JF}) does not yet determine the sign of $\bar{\epsilon}_{0}$. If $\bar{\epsilon}_{0} < 0$ we get deceleration or super-inflation in the Jordan frame \cite{Domenech:2015qoa}. However, it follows from (\ref{Friedmann2.EF}) that these phenomena are not possible in the Einstein frame. Secondly, notice that in the Einstein frame to obtain $|\hat{\epsilon}_{0}| \ll 1$ requires $|\hat{\kappa}_{0}|\ll 1$, whereas in the Jordan frame to obtain $|\bar{\epsilon}_{0}| \ll 1$ does not necessarily require the conditions (\ref{slow.roll.JF}) to be satisfied. This can be seen from (\ref{Friedmann2.JF1}), where it is at least theoretically possible that the special combination on the r.h.s.\ makes the l.h.s.\ small compared to each individual term on the r.h.s. Similarly, in the Einstein frame the approximations (\ref{slow.roll.EF.approximation}) require the conditions (\ref{slow.roll.EF}) to be satisfied, whereas in the Jordan frame it could be the case that the approximations (\ref{slow.roll.JF.approximation}) hold, but the conditions (\ref{slow.roll.JF}) do not.

For the leading order slow-roll approximation in the Einstein frame we introduced two slow-roll conditions (\ref{slow.roll.EF}), whereas in the Jordan frame we introduced four slow-roll conditions  (\ref{slow.roll.JF}). Let us investigate whether the conditions in the Jordan frame (\ref{slow.roll.JF}) imply the conditions in the Einstein frame (\ref{slow.roll.EF}). After a straightforward calculation we can express the Einstein frame parameters in terms of the Jordan frame parameters and vice versa:
%
\begin{subequations} 
%
\begin{eqnarray}
\label{kappa.EF.in.terms.of.JF}
\hat{\kappa}_{0} &=& \frac{ \bar{\kappa}_{0}}{ \left( 1- \bar{\lambda}_{0} \right)^2 }
\,, \qquad \qquad
\hat{\kappa}_{1} = \frac{\bar{\kappa}_{1}}{ 1 - \bar{\lambda}_{0} }
+\frac{ 2 \bar{\lambda}_{0} \bar{\lambda}_{1} }{ \left( 1- \bar{\lambda}_{0} \right)^2 } \,, \\
\bar{\kappa}_{0} &=& \frac{ \hat{\kappa}_{0}}{ \left( 1+ \hat{\lambda}_{0} \right)^2 } \,, \qquad \qquad
\bar{\kappa}_{1} = \frac{\hat{\kappa}_{1}}{ 1 + \hat{\lambda}_{0} }
- \frac{ 2 \hat{\lambda}_{0} \hat{\lambda}_{1} }{ \left( 1 + \hat{\lambda}_{0} \right)^2 }
\,.
\end{eqnarray}
%
\end{subequations}
From these expressions we deduce that the slow-roll conditions in the Jordan frame (\ref{slow.roll.JF}) imply the slow-roll conditions in the Einstein frame (\ref{slow.roll.EF}). The converse does not hold: we cannot deduce from the two conditions (\ref{slow.roll.EF}) that the four conditions (\ref{slow.roll.JF}) hold. However, we have defined parameters $\hat{\lambda}_{0}$ and $\hat{\lambda}_{1}$ (\ref{lambda.0.EF})-(\ref{lambda.1.EF}) similarly to (\ref{lambda.0.JF})-(\ref{lambda.1.JF}),
and a straightforward calculation yields that these parameters are related by
\vspace{-0.2cm}
\begin{equation}
\label{lambda.EF.in.terms.of.JF}
\hat{\lambda}_{0} = \frac{ \bar{\lambda}_{0} }{ 1 - \bar{\lambda}_{0} }\, , \qquad
\hat{\lambda}_{1} = \frac{ \bar{\lambda}_{1} }{ \left( 1 - \bar{\lambda}_{0} \right)^2 }
\,, \qquad
\bar{\lambda}_{0} = \frac{ \hat{\lambda}_{0} }{ 1 + \hat{\lambda}_{0} } \,, \qquad
\bar{\lambda}_{1} = \frac{ \hat{\lambda}_{1} }{ \left( 1 + \hat{\lambda}_{0} \right)^2 } \,.
\end{equation}
If additionally to the two slow-roll conditions (\ref{slow.roll.EF}) in the Einstein frame the conditions
(\ref{slow.roll.EF1}) are satisfied, then the slow-roll conditions (\ref{slow.roll.JF}) in the Jordan frame follow. Conversely, if the slow-roll conditions (\ref{slow.roll.JF}) in the Jordan frame hold, we have (\ref{slow.roll.EF}) and (\ref{slow.roll.EF1}) satisfied.

The discussion above can be understood more easily, if we explicitly write down how the parameters $\bar{\epsilon}_{0}$ and  $\hat{\epsilon}_{0}$, which define the inflationary regimes in different frames, are related:
\begin{equation}
\label{epsilon.0.JF.in.terms.of.EF}
\bar{\epsilon}_{0} = \frac{ \hat{\epsilon}_{0} + \hat{\lambda}_{0} }{  1  + \hat{\lambda}_{0} } -  \frac{   \hat{\lambda}_{0} \hat{\lambda}_{1}  }{ \left(  1  + \hat{\lambda}_{0} \right)^2 }
\,, \qquad \qquad
\hat{\epsilon}_{0} = \frac{  \bar{\epsilon}_{0} - \bar{\lambda}_{0} }{  1 - \bar{\lambda}_{0} } + \frac{ \bar{\lambda}_{0} \bar{\lambda}_{1} }{ \left( 1 - \bar{\lambda}_{0} \right)^{2} } \,.
\end{equation}
The slow-roll in the Einstein frame yields $\hat{\epsilon}_{0} \ll 1$, and it follows from the last expression that in order to obtain  $|\bar{\epsilon}_{0}| \ll 1$ we need  to further impose (\ref{slow.roll.EF1}), which, however, does not generally guarantee that $|\bar{\epsilon}_{0}|$ and  $\hat{\epsilon}_{0}$ are of the same order of magnitude. From (\ref{epsilon.0.JF.in.terms.of.EF}) it is clear that the condition $\mathcal{O}(|\bar{\epsilon}_{0}|)=\mathcal{O}(\hat{\epsilon}_{0})$ holds, if one assumes that $\hat{\epsilon}_{0}$ and $\hat{\lambda}_{0}$ are of the same order of magnitude. Due to the equality in (\ref{Friedmann2.EF1}) this assumption just yields that $\hat{\kappa}_{0}$ and $\hat{\lambda}_{0}$ are of the same order of magnitude. These parameters are related as $ \hat{\lambda}_{0}^2 = \mathcal{I}_5 \hat{\kappa}_0 $,
where the invariant $\mathcal{I}_5$ defined in (\ref{I.5}) represents the strength of the non-minimal coupling between the scalar field and gravity. In the case $\mathcal{I}_5=0$  the scalar field is minimally coupled to gravity and  we have $\hat{\lambda}_{0}=0$, which, if inserted into (\ref{epsilon.0.JF.in.terms.of.EF}), will yield $\bar{\epsilon}_{0}=\hat{\epsilon}_{0}$ as expected. By definition $\mathcal{I}_5$ is nonnegative.

It is important to notice that in some parametrizations $\mathcal{I}_5$ is bounded from above. As an example, if we consider the Jordan frame BEPS parametrization (see appendix \ref{App:Invariants.in.different.parametrizations}) and require that gravity is attractive, then $\mathcal{I}_5$ cannot exceed the value of  $\frac{1}{3}$. If we restrict the general theory to this parametrization we see that $\hat{\kappa}_0 \ll 1$ implies $\hat{\lambda}_{0} \ll 1$, however, we have generally $\mathcal{O}(\hat{\lambda}_{0}) \neq \mathcal{O}(\hat{\kappa}_{0})$. Thus in the extreme case with $\mathcal{I}_5=\frac{1}{3}$  one obtains  $\mathcal{O}(\hat{\epsilon}_{0}) = \mathcal{O}(\hat{\kappa}_{0})$, whereas $\mathcal{O}(\bar{\epsilon}_{0}) = \mathcal{O}(\sqrt{\hat{\kappa}_{0}})$. We can summarize the discussion as follows: in order to obtain $\bar{\epsilon}_{0}  \approx \hat{\epsilon}_{0}$ we need to assume that the parameters $\hat{\kappa}_{0}$ and $\hat{\lambda}_{0}$ are of the same order of magnitude. This is equivalent to the condition that $\bar{\kappa}_{0}$ and $\bar{\lambda}_{0}$ are of the same order of magnitude. Now consider the second slow-roll parameter in the Einstein frame, namely $\hat{\kappa}_{1}$. If we further assume that $\hat{\kappa}_{1}$ and $\hat{\lambda}_{0}$ are of the same order then we obtain that 
$ \bar{\kappa}_{1}\approx \hat{\kappa}_{1} - 2\hat{\lambda}_{0}\hat{\lambda}_{1}$. We can conclude that assuming  $\mathcal{O}(\hat{\lambda}_{0}) =\mathcal{O}(\hat{\kappa}_{0}) =\mathcal{O}(\hat{\kappa}_{1})= \mathcal{O}(\hat{\lambda}_{1}) $ is equivalent to $\mathcal{O}(\bar{\lambda}_{0}) =\mathcal{O}(\bar{\kappa}_{0}) =\mathcal{O}(\bar{\kappa}_{1})= \mathcal{O}(\bar{\lambda}_{1})$. These conditions also imply that $\mathcal{O}(\bar{\kappa}_{0}) =\mathcal{O}(\hat{\kappa}_{0})$. Similar assumptions were made in
\cite{Chiba:2008ia} when expressing the extended slow-roll parameters formulated in the Jordan frame BEPS in terms of the standard slow-roll parameters.

We have discussed the relation between $\bar{\epsilon}_{0}$ and $\hat{\epsilon}_{0}$, which are the HSR parameters. The slow-roll regime is sustained if these parameters remain small during the time scale given by the Hubble parameter. That motivates one to define the parameters
\begin{equation}
\label{higher.epsilon}
\bar{\epsilon}_{i} = \frac{1}{\bar{H} \bar{\epsilon}_{i-1} } \frac{ \mathrm{d} \bar{\epsilon}_{i-1} }{ \mathrm{d} \bar{t} } \, , \qquad \qquad
\hat{\epsilon}_{i} = \frac{1}{\hat{H} \hat{\epsilon}_{i-1} } \frac{ \mathrm{d} \hat{\epsilon}_{i-1} }{ \mathrm{d} \hat{t} } \,, \qquad \qquad i = 1,\, 2,\, 3\,, \ldots \,.
\end{equation}
Demanding that the parameters $\hat{\epsilon}_{i}$ remain small is essentially the standard slow-roll scenario in terms of derivatives of the Hubble parameter. Notice that $\hat{\epsilon}_{0}=\hat{\kappa}_{0}$ and $\hat{\epsilon}_{1}=\hat{\kappa}_{1}$, which illustrates that our formulation of the slow-roll regime in the Einstein frame (\ref{slow.roll.EF}) amounts to considering slow-roll in terms of derivatives of the Hubble parameter. Considering the slow-roll conditions in the Jordan frame (\ref{slow.roll.JF}) we see that they imply the smallness of $\bar{\epsilon}_{0}$, however, it is unclear if they imply the smallness of $\bar{\epsilon}_{1}$ as well. Let us explore how the parameters  $\bar{\epsilon}_{1}$ and $\hat{\epsilon}_{1}$ are related. Using (\ref{epsilon.0.JF.in.terms.of.EF}) in a straightforward calculation yields
\begin{equation}
\label{epsilon.1.JF.in.terms.of.EF}
\bar{\epsilon}_{1} = \frac{ \hat{\epsilon}_{1} + \hat{\lambda}_{1} \left\lbrace - \frac{ \hat{\lambda}_{0} }{ 1 +  \hat{\lambda}_{0} } + \frac{ \hat{\lambda}_{0} }{ \hat{\epsilon}_{0} }
\left[ 1 - \frac{1}{ ( 1 +  \hat{\lambda}_{0} )} \left( \hat{\lambda}_{0} + \hat{\lambda}_{1} + \hat{\lambda}_{2} - \frac{ 2 \hat{\lambda}_{0} \hat{\lambda}_{1} }{ 1 +  \hat{\lambda}_{0} } \right) \right] \right\rbrace }{ \left[ 1 +  \hat{\lambda}_{0} \right] \left[ 1 + \frac{ \hat{\lambda}_{0} }{ \hat{\epsilon}_{0} } \left( 1 - \frac{ \hat{\lambda}_{1} }{ 1 + \hat{\lambda}_{0} } \right) \right] } \,.
\end{equation}
Now, imposing  $\mathcal{O}(\hat{\lambda}_{0}) =\mathcal{O}(\hat{\kappa}_{0}) =\mathcal{O}(\hat{\kappa}_{1})= \mathcal{O}(\hat{\lambda}_{1})$, as well as that $\hat{\lambda}_2$ is of order unity or higher order small, will indeed yield $\mathcal{O}(\hat{\epsilon}_{1}) =\mathcal{O}(\bar{\epsilon}_{1})$.


\subsection{The role of matter}

While deriving the conditions for slow-roll inflation in different frames  we have completely neglected the matter fields. This assumption seems reasonable since the influence of matter is negligible in the inflationary regime. Recall that as mentioned after the system (\ref{Friedmann1.EF})-(\ref{Friedmann3.EF}), from the viewpoint of transformation properties neglecting matter is a covariant procedure \cite{Jarv:2015kga}. 

The general action functional (\ref{fl.moju}) contains four arbitrary functions of the scalar field and one of them, $\alpha$, represents the non-minimal coupling between the scalar field and the matter fields. So neglecting matter may discard one of the four functional degrees of freedom, in particular, if we fix the parametrization by leaving $\alpha$ arbitrary and then assume that matter is negligible. 

The situation where we first fix $\alpha = 0$ and then assume that matter is negligible, corresponds to the Jordan frame without matter, whereas the situation where we first fix the parametrization by leaving $\alpha$ arbitrary and then assume that matter is negligible, corresponds to the Einstein frame without matter. Thus we obtain that in the inflationary regime in the Jordan frame we have two functional degrees of freedom, whereas in the Einstein frame we have only one functional degree of freedom. Inflation in the Einstein frame can thus be described solely by the potential, whereas in the Jordan frame one has besides the potential an additional degree of freedom: a non-minimal coupling to gravity.

Now we can clarify why the slow-roll regime in the Einstein frame does not imply the slow-roll regime in the Jordan frame. By neglecting matter in the Einstein frame we hide one functional degree of freedom and cannot transform the theory to the Jordan frame, since we have deleted the information about the coupling of the scalar field and the matter. Due to this argument, one cannot deduce from the slow-roll conditions in the Einstein frame the slow-roll conditions in the Jordan frame since the latter is unspecified. However, formulating the inflationary regime in the Jordan frame and neglecting matter does not induce problems in transforming the theory to the Einstein frame. This explains why we obtained that the slow-roll regime in the Jordan frame implies the slow-roll regime in the Einstein frame but not vice versa.


\section{Spectral indices}\label{Sec:Spectral.indices}

The main motivation to consider slow-roll inflation is its ability to produce the curvature perturbation spectrum that seeds the structure formation and is measurable through the cosmic microwave background radiation. Working in the Einstein frame, one can readily use the expressions for the spectral indices calculated in the minimally coupled theory. Similar calculations have been carried out also in the Jordan frame \cite{Chiba:2008ia,Kaiser:1995nv,Noh:2001ia}, giving identical results as in the Einstein frame, up to first order in slow-roll parameters. The apparent discrepancy between different frames already in first order arose probably due to the frame-dependent definition of the curvature perturbation used in earlier works \cite{Kaiser:1994vs,Kaiser:1995nv}. In \cite{Nozari:2010uu}, a difference in second order was found. Here we repeat the calculation by making use of the scheme put forward in \cite{Stewart:1993bc} and conclude that due to the underlying assumptions, the results in both frames coincide. 


\subsection{Invariant form of the Mukhanov-Sasaki equation}

The Mukhanov-Sasaki equation describes the behaviour of linear curvature perturbations in a flat FLRW universe filled with a scalar field \cite{Mukhanov.Sasaki}. Following \cite{vandeBruck:2015gjd,Hwang:2005hb}, it reads
\begin{equation}
\label{mukhanov.sasaki}
\frac{ \mathrm{d}^2 v }{ \mathrm{d} \tau^2 } + \left( k^2 -\frac{ 1 }{z} \frac{ \mathrm{d}^2 z }{ \mathrm{d} \tau^2 } \right) v = 0 \,,
\end{equation}
where $\tau$ is the conformal time; $k$ is the comoving wave number of the gauge invariant curvature perturbation $\mathcal{R}$; $v \equiv z \mathcal{R} \equiv a \sqrt{Q} \, \mathcal{R}$, where $a$ is the scale factor and
\begin{equation}
\label{Q}
Q \equiv \frac{ 2\mathcal{B} \dot{\Phi}^2 + 3 \frac{ \dot{\mathcal{A}}^2 }{ \mathcal{A} } }{ 2 \left( H + \frac{ \dot{\mathcal{A}} }{ 2\mathcal{A} } \right)^2 } = \frac{2\mathcal{F} \dot{\Phi}^2}{\hat{H}^2}= 2 \mathcal{A} \hat{\kappa}_{0}
\,.
\end{equation}
Note that the definition of $Q$ can be used in each parametrization, upon suitably specifying $\mathcal{A}$ and $\mathcal{B}$.

Here we shall confirm explicitly the parametrization independence of the Mukhanov-Sasaki equation (\ref{mukhanov.sasaki}) discussed also in \cite{Domenech:2015qoa}. The comoving wave number $k$ does not depend on the parametrization \cite{Hwang:2005hb} as well as the gauge invariant curvature perturbation $\mathcal{R}$ \cite{Chiba:2008ia,Gong:2011qe}. In order for (\ref{mukhanov.sasaki}) to be parametrization independent $z$ should be an invariant. This is indeed the case since we can express
\begin{equation}
\label{z.invariant}
z = a \sqrt{Q} = \sqrt{2} a \dot{\Phi} \frac{ \sqrt{\mathcal{F}} }{ \hat{H} } = \pm \sqrt{2} \frac{ \hat{a} }{ \hat{H} } \frac{ \mathrm{d} \mathcal{I}_3 }{ \mathrm{d} \hat{t} } = \pm \sqrt{\frac{2}{\mathcal{I}_{1}}} \frac{\bar{a} }{ \bar{H} (1-\bar{\lambda}_{0})}\frac{ \mathrm{d} \mathcal{I}_3 }{ \mathrm{d} \bar{t} } \,,
\end{equation}
which is an invariant. Thus we conclude that the equation (\ref{mukhanov.sasaki}) is parametrization independent. This fact is compatible with the presumption that the spectral indices, calculated via the solutions of the Mukhanov-Sasaki equation, should not depend on the parametrization.


\subsection{Calculating scalar spectral index}


\subsubsection{Einstein frame}

Let us perform the calculation of the scalar spectral index in the Einstein frame following \cite{Stewart:1993bc}. The term $z^{-1}  \mathrm{d}^2 z / \mathrm{d} \tau^2$ in the Mukhanov-Sasaki equation (\ref{mukhanov.sasaki}) can be expanded in terms of the slow-roll parameters:
\begin{equation}
\label{z.EF}
\frac{1}{z} \frac{ \mathrm{d}^2 z }{ \mathrm{d} \tau^2 } = 2 \hat{a}^2 \hat{H}^2 \left( 1 - \frac{1}{2} \hat{\epsilon}_{0} + \frac{3}{4} \hat{\kappa}_{1} + \frac{1}{8} \hat{\kappa}_{1}^{2} - \frac{1}{4}  \hat{\epsilon}_{0} \hat{\kappa}_{1} + \frac{1}{4} \hat{\kappa}_{1} \hat{\kappa}_{2} \right)
\,.
\end{equation}
The conformal time is given by
\begin{equation}
\label{conformal.time.EF}
\tau = \int \frac{\mathrm{d} \hat{t}}{\hat{a}} = - \frac{1}{ \hat{a} \hat{H} } + \int \hat{\epsilon}_0 \frac{ \mathrm{d} \hat{a} }{ \hat{a}^2 \hat{H} }
\,.
\end{equation}
Assuming that $\hat{\epsilon}_{0}$, $\hat{\kappa}_{1}$ and $\hat{\kappa}_{2}$ are small and change negligibly during $\hat{H}^{-1}$, one can treat the expression inside the brackets of (\ref{z.EF}) as a constant. Taking  $\hat{\epsilon}_{0}$ as a constant, the conformal time is given by
\begin{equation}
\label{conformal.time.EF1}
\tau \approx - \frac{1}{ \hat{a} \hat{H} } \left( \frac{1}{ 1 - \hat{\epsilon}_{0} } \right)
\end{equation}
and the expression (\ref{z.EF}) reduces to
\begin{equation}
\label{z.EF1}
\frac{1}{z} \frac{ \mathrm{d}^2 z }{ \mathrm{d} \tau^2 } \approx \frac{1}{\tau^2} \left( \nu^2 - \frac{1}{4} \right)
\,,
\end{equation}
where the parameter $\nu$ reads
\begin{equation} 
\label{nu.EF}
\nu = \left\lbrace \frac{2}{ ( 1 - \hat{\epsilon}_{0} )^2 } \left( 1 - \frac{1}{2} \hat{\epsilon}_{0} + \frac{3}{4} \hat{\kappa}_{1} + \frac{1}{8} \hat{\kappa}_{1}^{2} - \frac{1}{4} \hat{\epsilon}_{0} \hat{\kappa}_{1} + \frac{1}{4} \hat{\kappa}_{1} \hat{\kappa}_{2} \right) + \frac{1}{4} \right\rbrace^{ \frac{1}{2} }
\,.
\end{equation}
Under these assumptions the solution of (\ref{mukhanov.sasaki}) is given by \cite{Stewart:1993bc}
\begin{equation}
\label{v.MS}
v = \frac{1}{2} \sqrt{ \pi } \, \mathrm{exp}\left[ i \left( \nu + \frac{1}{2} \right) \frac{ \pi }{2} \right] (-\tau )^{ \frac{1}{2} } H_{\nu}^{\, (1)}( -k\tau )
\,,
\end{equation}
and the square root of the spectrum for the curvature perturbation at the horizon crossing $k=\hat{a}\hat{H}$ reads
\begin{equation}
\label{spectrum.EF}
\hat{P}_{\mathcal{R}}^{\, \frac{1}{2} }(k) = \left. 2^{ \nu - \frac{3}{2} } \frac{ \Gamma(\nu) }{ \Gamma \left( \frac{3}{2} \right) }
( 1 - \hat{\epsilon}_0 )^{ \nu - \frac{1}{2} } \frac{ \hat{H}^2 }{ 2 \sqrt{2} \pi } \left|  \frac{ \mathrm{d} \mathcal{I}_3 }{ \mathrm{d} \hat{t} } \right|^{-1} \right|_{ \hat{a} \hat{H} = k } 
\,.
\end{equation}
We have from (\ref{nu.EF}) to first order in slow-roll parameters 
\begin{equation}
\label{nu.EF1}
\nu \approx \frac{3}{2} + \hat{\epsilon}_{0} + \frac{1}{2} \hat{\kappa}_{1}
\, 
\end{equation}
and (\ref{spectrum.EF}) can be approximated
\begin{equation}
\label{spectrum.EF1}
\hat{P}_{\mathcal{R}}^{\, \frac{1}{2}}(k) \approx \left.
\left[ 1 + ( 2-\ln 2-b )( \hat{\epsilon}_{0} + \frac{1}{2} \hat{\kappa}_{1} ) - \hat{\epsilon}_{0} \right] \frac{ \hat{H}^2 }{ 2 \sqrt{2} \pi } \left| \frac{ \mathrm{d} \mathcal{I}_3 }{ \mathrm{d} \hat{t} } \right|^{-1} \right|_{ \hat{a}\hat{H} = k } 
\,,
\end{equation}
where $b$ is the Euler-Mascheroni constant. The spectral index $\hat{n}_{_\mathrm{S}}$ is obtained by taking the logarithmic derivative of the spectrum at the horizon crossing:
\begin{equation}
\label{n.EF}
\hat{n}_{_\mathrm{S}} - 1 \equiv  \left. 2 \frac{ \mathrm{d} \ln \hat{P}_{ \mathcal{R} }^{\, \frac{1}{2}}(k) }{ \mathrm{d} \ln k } \right|_{ \hat{a}\hat{H} = k } 
\,.
\end{equation}
A direct calculation up to second order in the slow-roll parameters yields
\begin{equation}
\label{n.EF1}
\hat{n}_{_\mathrm{S}} - 1 = - 2 \hat{\kappa}_{0} - \hat{\kappa}_{1} - 2 \hat{\kappa}_{0}^{\, 2} + ( 1 - 2 \ln 2-2b ) \hat{\kappa}_{0} \hat{\kappa}_{1} + ( 2 - \ln 2-b ) \hat{\kappa}_{1} \hat{\kappa}_{2}
\,.
\end{equation}


\subsubsection{Jordan frame}

Let us perform the calculation of the spectral index in the Jordan frame assuming slow-roll (\ref{slow.roll.JF}) in the Jordan frame. The term $z^{-1} \mathrm{d}^2 z / \mathrm{d} \tau^2$ in the Mukhanov-Sasaki equation  (\ref{mukhanov.sasaki}) can be expanded in terms of the slow-roll parameters:
\begin{equation}
\label{z.JF}
\frac{1}{z} \frac{ \mathrm{d}^2 z }{ \mathrm{d} \tau^2 } = 2   \bar{a}^2 \bar{H}^2 \left( 1 - \frac{1}{2} \bar{\epsilon}_{0} + \frac{3}{4} \bar{\kappa}_{1} - \frac{3}{2} \bar{\lambda}_{0} + \dots \right)
\,.
\end{equation}
Here we have omitted the terms of second order in slow-roll parameters. Note that (\ref{z.JF}) and (\ref{z.EF}) are the same expressions just written using different variables. The conformal time is given by
\begin{equation}
\label{conformal.time.JF}
\tau = \int \frac{ \mathrm{d} \bar{t} }{ \bar{a} } = - \frac{1}{ \bar{a} \bar{H} } + \int \bar{\epsilon}_0 \frac{ \mathrm{d} \bar{a} }{ \bar{a}^2 \bar{H} }
\, .
\end{equation}
Now assuming that $\bar{\epsilon}_{0}$, $\bar{\kappa}_{1}$, $\bar{\kappa}_{2}$, $\bar{\lambda}_{0}$, $\bar{\lambda}_{1}$ and $\bar{\lambda}_{2}$ are small and change negligibly during $\bar{H}^{-1}$, one can treat the expression inside the brackets of (\ref{z.JF}) as a constant. Taking  $\bar{\epsilon}_{0}$ as a constant, the conformal time reads
\begin{equation}
\label{conformal.time.JF1}
\tau \approx - \frac{1}{ \bar{a}\bar{H} } \left( \frac{1}{ 1 - \bar{\epsilon}_{0} } \right)
\end{equation}
and the expression (\ref{z.JF}) reduces to
\begin{equation}
\label{z.JF1}
\frac{1}{z} \frac{ \mathrm{d}^2 z }{ \mathrm{d} \tau^2 } \approx \frac{1}{ \tau^2 } \left( \nu^2 - \frac{1}{4} \right)
\,,
\end{equation}
where the parameter $\nu$ up to first order in slow-roll parameters is given by
\begin{equation}
\label{nu.JF}
\nu \approx \frac{3}{2} + \bar{\epsilon}_{0} + \frac{1}{2} \bar{\kappa}_{1} - \bar{\lambda}_{0}
\,.
\end{equation}
Our ansatz for solving the Mukhanov-Sasaki equation is similar as in the case of the calculation in the Einstein frame, thus the solution of (\ref{mukhanov.sasaki}) is given by (\ref{v.MS}). If we calculate the square root of the spectrum for the curvature perturbation at the horizon crossing in the Jordan frame $k=\bar{a}\bar{H}$ we obtain
\begin{equation}
\label{spectrum.JF}
\bar{P}_{\mathcal{R}}^{\, \frac{1}{2}}(k) = \left. 2^{ \nu - \frac{3}{2} } \frac{ \Gamma(\nu) }{ \Gamma(\frac{3}{2}) }
( 1 - \bar{\epsilon}_0 )^{ \nu - \frac{1}{2} } \frac{ \bar{H}^2 }{ 2 \sqrt{2} \pi } \left| \frac{ \mathrm{d} \mathcal{I}_3 }{ \mathrm{d} \bar{t} } \right|^{-1} \sqrt{ \mathcal{I}_1 }( 1 - \bar{\lambda}_0 ) \right|_{ \bar{a}\bar{H} = k } 
\,.
\end{equation}
Using (\ref{nu.JF}) we can approximate
\begin{equation}
\label{spectrum.JF1}
\bar{P}_{\mathcal{R}}^{\, \frac{1}{2}}(k) \approx \left.
\left[ 1 + ( 2 - \ln 2 - b )( \bar{\epsilon}_{0} - \bar{\lambda}_{0} +  \frac{1}{2} \bar{\kappa}_{1} ) - \bar{\epsilon}_{0} \right] \frac{ \bar{H}^2 }{ 2 \sqrt{2} \pi } \left| \frac{ \mathrm{d} \mathcal{I}_3 }{ \mathrm{d} \bar{t} } \right|^{-1} \sqrt{ \mathcal{I}_1 }( 1 - \bar{\lambda}_0 ) \right|_{ \bar{a}\bar{H} = k }
\,.
\end{equation}
The spectral index in the Jordan frame $\bar{n}_{_\mathrm{S}}$ is obtained by taking the logarithmic derivative of the spectrum at the horizon crossing:
\begin{equation}
\label{n.JF}
\bar{n}_{_\mathrm{S}} - 1 \equiv  \left. 2 \frac{ \mathrm{d} \ln \bar{P}_{\mathcal{R}}^{\, \frac{1}{2}}(k) }{ \mathrm{d} \ln k } \right|_{ \bar{a}\bar{H} = k } 
\, .
\end{equation}
A direct calculation up to second order in slow-roll parameters yields
\begin{eqnarray}
\label{n.JF1}
\nonumber 
\bar{n}_{_\mathrm{S}} - 1 &=& - 2 \bar{\kappa}_{0} - \bar{\kappa}_{1} - 2 \bar{\kappa}_{0}^{\, 2} + ( 1 - 2 \ln 2 - 2b ) \bar{\kappa}_{0} \bar{\kappa}_{1} + ( 2 - \ln 2 - b ) \bar{\kappa}_{1} \bar{\kappa}_{2} 
\\
&&- \bar{\lambda}_{0} \left( 4 \bar{\kappa}_{0} + \bar{\kappa}_{1} + 2\bar{\lambda}_{1} \right)
\,.
\end{eqnarray}

Let us compare this result with the calculation in the Einstein frame (\ref{n.EF1}). Assuming the slow-roll regime in both frames, we can express the Einstein frame parameters in terms of the Jordan frame parameters via (\ref{kappa.EF.in.terms.of.JF}) as
\begin{equation}
\label{parameters1}
\hat{\kappa}_{0} \approx \bar{\kappa}_{0} + 2 \bar{\kappa}_{0} \bar{\lambda}_{0} \,, \qquad \qquad
\hat{\kappa}_{1} \approx \bar{\kappa}_{1} + \bar{\kappa}_{1} \bar{\lambda}_{0} + 2 \bar{\lambda}_{0} \bar{\lambda}_{1} \,, \qquad \qquad
\hat{\kappa}_{2} \approx \bar{\kappa}_{2}
\,,
\end{equation}
where the first two approximations are up to second order and the third up to first order in slow-roll parameters. Plugging these expressions into (\ref{n.EF1}) we obtain
\begin{equation}
\label{n.difference}
\hat{n}_{_\mathrm{S}} \approx \bar{n}_{_\mathrm{S}}
\,.
\end{equation}
Thus the scalar spectral indices calculated in different frames up to  second order in slow-roll parameters coincide. Note that at a fixed moment the comoving wave number $k$ at the horizon crossing is slightly different in different frames, $\bar{a}\bar{H} = \hat{a}\hat{H} \left(1 + \hat{\lambda}_{0}  \right)$. We have neglected this issue as it should not affect the second order results (\ref{n.difference}).


\subsection{Calculating tensor spectral index}


\subsubsection{Einstein frame}

The differential equation governing tensor perturbations can be recast in the Mukhanov-Sasaki form (\ref{mukhanov.sasaki}), where now $z \equiv \hat{a}$, $v \equiv z \psi$ and $\psi$ denotes the mode function for tensor perturbations. It is easy to see that also in this case the form of the Mukhanov-Sasaki equation is preserved under the change of parametrization.  The calculation of the spectral index for tensor modes $\hat{n}_{_\mathrm{T}}$ is similar to the previous calculation of the spectral index for the curvature perturbation \cite{Stewart:1993bc} while the definition for $\hat{n}_{_\mathrm{T}}$ is taken from \cite{Lyth:2009zz}. Firstly, by assuming $\hat{\epsilon}_{0}$ to be approximately constant we get the expression (\ref{conformal.time.EF1}) for the conformal time. Next calculate the term $z^{-1} \mathrm{d}^2 z / \mathrm{d} \tau^2$:
\begin{equation}
\label{z.EF2}
\frac{1}{z} \frac{ \mathrm{d}^2 z }{ \mathrm{d} \tau^2 } = 2 \hat{a}^2 \hat{H}^2 \left( 1 - \frac{1}{2} \hat{\epsilon}_{0} \right)
\approx \frac{2}{\tau^2 ( 1 - \hat{\epsilon}_{0} )^2} \left( 1 - \frac{1}{2} \hat{\epsilon}_{0} \right) \,, \qquad z \equiv \hat{a} = \frac{\bar{a}}{ \sqrt{\mathcal{I}_{1}} }
\,,
\end{equation}
where for the approximation we used (\ref{conformal.time.EF1}). By assuming $\hat{\epsilon}_{0}$ to be approximately constant, we can recast (\ref{z.EF2}) in the form of (\ref{z.EF1}),
where the parameter $\nu$ up to first order in slow-roll parameters reduces to
\begin{equation}
\label{nu.EF2}
\nu \approx \frac{3}{2} + \hat{\epsilon}_{0}
\,.
\end{equation}
Under these assumptions the solution of (\ref{mukhanov.sasaki}) is given by (\ref{v.MS}), from which one  deduces that the the square root of the spectrum for the tensor modes at the horizon crossing $k=\hat{a}\hat{H}$ reads
\begin{equation}
\label{spectrum.EF2}
\hat{P}_{\psi}^{\, \frac{1}{2}}(k) = \left. 2^{ \nu -\frac{3}{2} } \frac{ \Gamma(\nu) }{ \Gamma( \frac{3}{2} ) }
( 1 - \hat{\epsilon}_0 )^{ \nu - \frac{1}{2} } \frac{ \hat{H} }{ 2 \pi } \right|_{ \hat{a}\hat{H} = k } 
\, .
\end{equation}
Now using (\ref{nu.EF2}) yields
\begin{equation}
\label{spectrum.EF3}
\hat{P}_{\psi}^{\, \frac{1}{2}}(k) \approx  \left.
\left[ 1 + ( 1- \ln 2 - b ) \hat{\epsilon}_{0} \right] \frac{ \hat{H} }{ 2 \pi } \right|_{ \hat{a}\hat{H} = k } 
\, ,
\end{equation}
where $b$ is the Euler-Mascheroni constant. 
The spectral index $\hat{n}_{_\mathrm{T}}$ is obtained by taking the logarithmic derivative of the spectrum at the horizon crossing:
\begin{equation}
\label{n.EF2}
\hat{n}_{_\mathrm{T}} \equiv \left. 2 \frac{ \mathrm{d} \ln \hat{P}_{\psi}^{\, \frac{1}{2}}(k) }{ \mathrm{d} \ln k } \right|_{ \hat{a}\hat{H} = k } 
\,.
\end{equation}
A direct calculation up to second order in slow-roll parameters yields
\begin{equation}
\label{n.EF3}
\hat{n}_{_\mathrm{T}} = -2 \hat{\kappa}_{0} - 2\hat{\kappa}_{0}^{\, 2} + 2 ( 1 - \ln 2 - b ) \hat{\kappa}_{0} \hat{\kappa}_{1}
\,.
\end{equation}


\subsubsection{Jordan frame}

The spectral index for tensor modes in the Jordan frame $\bar{n}_{_\mathrm{T}}$ can be obtained similarly to the previous calculation carried out in the Einstein frame. First we assume $\bar{\epsilon}_{0}$ to be approximately constant to get the expression (\ref{conformal.time.JF1}) for the conformal time. Next calculate the term $z^{-1} \mathrm{d}^2 z / \mathrm{d} \tau^2$:
\begin{eqnarray}
\label{z.JF2}
\nonumber 
\frac{1}{z} \frac{ \mathrm{d}^2 z }{ \mathrm{d} \tau^2 } &=& 2 \bar{a}^2 \bar{H}^2 \left( 1 - \frac{1}{2} \bar{\epsilon}_{0}
- \frac{1}{2} \bar{\lambda}_{0} - \frac{ \bar{\lambda}_{0} \bar{\lambda}_{1} }{ 2 ( 1 - \bar{\lambda}_{0} ) } \right)
\left( 1 - \bar{\lambda}_{0} \right)\\
&\approx& \frac{2}{\tau^2 ( 1 - \bar{\epsilon}_{0} )^2 } \left( 1 - \frac{1}{2} \bar{\epsilon}_{0} - \frac{1}{2} \bar{\lambda}_{0} - \frac{\bar{\lambda}_{0} \bar{\lambda}_{1} }{ 2 ( 1 - \bar{\lambda}_{0} ) } \right)
\left( 1 - \bar{\lambda}_{0} \right)
\, ,
\end{eqnarray}
where for the approximated equality we used (\ref{conformal.time.JF1}). By assuming $\bar{\epsilon}_{0}$ to be constant we can recast (\ref{z.JF2}) in the form of (\ref{z.JF1}),
where the parameter $\nu$ up to first order in slow-roll parameters is given by 
\begin{equation}
\label{nu.JF2}
\nu \approx \frac{3}{2} + \bar{\epsilon}_{0} - \bar{\lambda}_{0}
\,.
\end{equation}
Under the imposed assumptions the solution of (\ref{mukhanov.sasaki}) is given by (\ref{v.MS}).  It follows that the square root of the spectrum for the tensor modes at the horizon crossing $k=\bar{a}\bar{H}$ reads
\begin{equation}
\label{spectrum.JF2}
\bar{P}_{\psi}^{\, \frac{1}{2}}(k) = \left. 2^{ \nu - \frac{3}{2} } \frac{ \Gamma(\nu) }{ \Gamma( \frac{3}{2} ) } ( 1 - \bar{\epsilon}_0 )^{ \nu - \frac{1}{2} } \frac{ \bar{H} }{ 2 \pi } \sqrt{ \mathcal{I}_1 } \right|_{ \bar{a}\bar{H} = k } 
\,.
\end{equation}
By using (\ref{nu.JF2}) we can approximate
\begin{equation}
\label{spectrum.JF3}
\bar{P}_{\psi}^{\, \frac{1}{2}}(k) \approx \left.
\left[ 1 + ( 2 - \ln 2 - b )( \bar{\epsilon}_{0} - \bar{\lambda}_{0} ) - \bar{\epsilon}_{0} \right] \frac{ \bar{H} }{ 2 \pi } \sqrt{ \mathcal{I}_1 } \right|_{ \bar{a}\bar{H} = k } 
\,.
\end{equation}
The spectral index $\bar{n}_{_\mathrm{T}}$ is obtained by taking the logarithmic derivative of the spectrum at the horizon crossing:
\begin{equation}
\label{n.JF2}
\bar{n}_{_\mathrm{T}} \equiv \left. 2 \frac{ \mathrm{d} \ln \bar{P}_{\psi}^{\, \frac{1}{2}}(k) }{ \mathrm{d} \ln k } \right|_{ \bar{a}\bar{H} = k } 
\,.
\end{equation}
A direct calculation up to second order in slow-roll parameters yields
\begin{equation}
\label{n.JF3}
\bar{n}_{_\mathrm{T}} = -2 \bar{\kappa}_{0} - 2\bar{\kappa}_{0}^{\, 2} + 2 ( 1 - \ln 2 - b ) \bar{\kappa}_{0} \bar{\kappa}_{1}
- 4\bar{\kappa}_{0} \bar{\lambda}_{0}
\,.
\end{equation}
Let us compare the tensor indices obtained in the Einstein and Jordan frames. Using (\ref{parameters1}) in
(\ref{n.EF3}) we get 
\begin{equation}
\label{n.difference1}
\hat{n}_{_\mathrm{T}} \approx \bar{n}_{_\mathrm{T}}
\,.
\end{equation}
Thus also the tensor spectral indices calculated in different frames up to second order in slow-roll parameters coincide.


\subsection{Observables via invariant potential \texorpdfstring{$\mathcal{I}_{2}$}{I\_2}}

In order to express spectral indices (\ref{n.EF1}) and (\ref{n.EF3}) in terms of the potential as well, we make use of the fact that in the Einstein frame the equations are formally the same as for the minimally coupled case. Hence we can rely on the results of Liddle {\it et al} \cite{Liddle:1994dx}, where an exhaustive treatment of the relations between HSR and PSR parameters was given. Their definitions of the parameters $(\epsilon_{_H}$, $\eta_{_H}$, $\xi_{_H})$ slightly differ from ours. The correspondence is the following:
%
\begin{subequations}
	%
	\begin{eqnarray}
	\hat{\kappa}_{0} &=& \hat{\epsilon}_{0} = \epsilon_{_H} \,, \qquad \qquad \qquad \qquad
	\hat{\kappa}_{1} = 2 (-\hat{\eta} + \hat{\epsilon}_{0}) = 2( - \eta_{_H} + \epsilon_{_H} ) \,, \\
	\hat{\kappa}_{1} \hat{\kappa}_{2} &=& 2 \left( 2 \epsilon_{_H}^{2} - 3 \epsilon_{_H} \eta_{_H} + \xi_{_H}^{2} \right)
	\end{eqnarray}
	%
\end{subequations}
for HSR parameters. In addition, by making use of the invariant $\mathcal{I}_2$ (\ref{I.1.I.2}), characterizing the potential $\mathcal{V}$, let us define PSR in the Einstein frame analogously to \cite{Liddle:1994dx}, i.e.\
\begin{equation}
\label{PSR.for.I2}
 \epsilon_{_V} \equiv \frac{1}{ 4 \mathcal{I}_{2}^{2} } \left( \frac{ \mathrm{d} \mathcal{I}_{2} }{ \mathrm{d} \mathcal{I}_{3} } \right)^2 \equiv \hat{\kappa}^{(\mathcal{V})}_{0} \,, \qquad \qquad
 \eta_{_V} \equiv \frac{1}{ 2 \mathcal{I}_{2} } \frac{ \mathrm{d}^{2} \mathcal{I}_{2} }{ \mathrm{d} \mathcal{I}_{3}^{2} } \,, \qquad \qquad
 \xi_{_V}^{2} \equiv \frac{1}{ 4 \mathcal{I}_{2}^{2}} \frac{ \mathrm{d} \mathcal{I}_{2} }{ \mathrm{d} \mathcal{I}_{3} } \frac{ \mathrm{d}^3 \mathcal{I}_{2} }{ \mathrm{d} \mathcal{I}_{3}^{3} } \,.
\end{equation}
From \cite{Liddle:1994dx} we can write out the relations between $\hat{\kappa}_{i}$ and $\left\lbrace \epsilon_{_V},\, \eta_{_V},\, \xi^{2}_{_V} \right\rbrace$ up to second order as
\begin{subequations}
%
\begin{eqnarray}
\label{HSR.vr.PSR.so1}
\hat{\kappa}_{0} &\approx& \epsilon_{_V} - \frac{4}{3} \epsilon_{_V}^{2} + \frac{2}{3} \epsilon_{_V} \eta_{_V} \,,  \qquad 
\hat{\kappa}_{1} \approx 2 \left( 2 \epsilon_{_V} - \eta_{_V} - 4 \epsilon_{_V}^{2} + \frac{10}{3} \epsilon_{_V} \eta_{_V} - \frac{1}{3} \eta_{_V}^{2} - \frac{1}{3} \xi_{_V}^{2} \right) \,, \\
\label{HSR.vr.PSR.so2}
\hat{\kappa}_{1} \hat{\kappa}_{2} &\approx& 2 \left( 8 \epsilon_{_V}^{2} - 6 \epsilon_{_V} \eta_{_V} + \xi_{_V}^{2} \right) \,.
\end{eqnarray}
%
\end{subequations}
Let us point out that expressions (\ref{PSR.for.I2}) are manifestly invariant and may be used as definitions in the Jordan frame as well. 

For the scalar spectral index in the Einstein frame, given by (\ref{n.EF1}), we obtain
\begin{eqnarray}    \nonumber
\hat{n}_{_\mathrm{S}} - 1 &\approx& \, - 4 \epsilon_{_H} + 2 \eta_{_H} + 8 \epsilon_{_H}^{2} (1 - \ln 2 - b) - 2 \epsilon_{_H} \eta_{_H} (7 - 5 \ln 2 - 5 b) + 2 \xi_{_H}^{2} \left( 2 - \ln 2 - b \right) \\
\nonumber
&\approx& - 6 \epsilon_{_V} + 2 \eta_{_V} + 2 \epsilon_{_V}^{2} \left( 22 + \frac{1}{3} - 12 \ln 2 - 12 b \right) -2 \epsilon_{_V} \eta_{_V} \left( 17 -8 \ln 2 -8 b \right) + \frac{2}{3} \eta_{_V}^{2}  \\
\label{n.S.for.I2}
&&+ 2 \xi_{_V}^{2} \left(  2 + \frac{1}{3} - \ln 2 -  b \right) \,.
\end{eqnarray}
Analogously for the tensor spectral index, given by (\ref{n.EF3})
\begin{eqnarray}
\nonumber
\hat{n}_{_\mathrm{T}} &\approx& - 2\epsilon_{_H} + 2 \epsilon_{_H}^2 \left( 1 - 2 \ln 2 - 2 b \right) -  4 \epsilon_{_H} \eta_{_H} \left( 1 - \ln 2 - b \right) \\
\label{n.T.for.I2}
&\approx& - 2 \epsilon_{_V} + 2 \epsilon_{_V}^{2} \left( 4 + \frac{1}{3} - 4 \ln 2 - 4 b \right) - 2 \epsilon_{_V} \eta_{_V} \left(3 - \frac{1}{3} - 2 \ln 2 - 2 b \right)\,.
\end{eqnarray}

Previously we have shown (see equations (\ref{n.difference}) and (\ref{n.difference1})) that up to second order these expressions are invariant under the change of parametrization. Hence, in the Jordan frame also the expressions (\ref{n.S.for.I2}) and (\ref{n.T.for.I2}) hold.


\section{Summary}\label{Sec:Summary}

We studied the slow-roll conditions for inflation in the framework of the general scalar-tensor theory of gravitation with the action functional (\ref{fl.moju}) using quantities (\ref{I.1.I.2})-(\ref{I.5}), which are invariant under conformal transformation and transform as scalar functions under reparametrization of the scalar field. We introduced the Einstein and Jordan frames with invariant metrics $\hat{g}_{\mu\nu}$ (\ref{Invariant.metric1}) and $\bar{g}_{\mu\nu}$ (\ref{Invariant.metric2}) correspondingly. Since these frames are formally related by a conformal transformation (\ref{Invariant.metric2}), the mutual correspondence of their properties is
similar to the usual relation between familiar Einstein and
Jordan frames. We proposed that the measurable metric is the ordinary Jordan frame metric $g_{\mu\nu}$. We emphasized the usefulness of the invariant Jordan frame (see subsubsection~\ref{tensorial.invariants}), which encodes all familiar frames upon specifying two of the four functional degrees of freedom $\{ \mathcal{A},\, 
\mathcal{B},\, \mathcal{V},\, \alpha \}$, including the usual Einstein 
and Jordan frames in their familiar form.   

We defined the slow-roll parameters (\ref{kappa.EF}) in the invariant Einstein frame by analogy to standard slow-roll inflation in general relativity with a minimally coupled scalar field. By defining parameters (\ref{kappa.JF}), (\ref{lambda.0.JF})-(\ref{lambda.1.JF}) we demonstrated how the slow-roll regime in the invariant Jordan frame can be formulated in a manner that implies slow-roll inflation in the invariant Einstein frame as well, i.e.\ from (\ref{slow.roll.JF}) also (\ref{slow.roll.EF}) follows. However, the slow-roll regime in the Jordan frame contains additional slow-roll parameters, and the slow-roll regime in the Einstein frame does not generally imply the slow-roll regime in the Jordan frame. This follows from the fact that by neglecting matter in the Einstein frame ($\alpha \neq 0$) we discard $\alpha$, one of the functional degrees of freedom, while in the Jordan frame we have by definition fixed ${\alpha} \equiv 0$. We compared in detail the slow-roll parameters in both frames and noticed that in certain parametrizations the smallness of the slow-roll parameters in the Einstein frame imply the smallness of the slow-roll parameters in the Jordan frame.   

Following the scheme given by Stewart and Lyth \cite{Stewart:1993bc}, we calculated the scalar and tensor spectral indices in the Einstein ((\ref{n.EF1}), (\ref{n.EF3})) and in the Jordan ((\ref{n.JF1}), (\ref{n.JF3})) frame up to second order and confirmed that they coincide ((\ref{n.difference}) as well as (\ref{n.difference1})). 

As an outlook it would be interesting to investigate those properties of cosmological models which seem to be frame-dependent, e.g.\ consider conditions for the end of inflation. Suitable examples with concrete potentials would certainly clarify and justify the formalism of invariants we presented. Another question that is not fully solved, was raised in \cite{ArmendarizPicon:2015dma}. Namely, although the observables describing spectrum are frame-independent, the scale of inflation depends on the conformal frame, and hence in the case of non-minimal coupling, distinctly from the minimally coupled theory, the measurement of the tensor amplitude would not characterize the scale of inflation completely.


\bigskip

\section*{Acknowledgments}

This work was supported by the Estonian Research Council Grant No.\ IUT02-27 and Grant No.\ PUT790, by the European Regional Development Fund through the Centre of Excellence 2014-2020.4.01.15-0004 and the Centre of Excellence 3.2.0101.11-0029. The authors are grateful to the referee for valuable comments.


\appendix

\section*{Appendices}
\addcontentsline{toc}{section}{Appendices}
\renewcommand{\thesubsection}{\Alph{subsection}}

\setcounter{section}{1}

\setcounter{equation}{0}
\setcounter{table}{0}
\numberwithin{equation}{subsection}
\numberwithin{table}{subsection}


\subsection{Invariants in different parametrizations \texorpdfstring{\cite{JKSV.Invariants}}{}} 
\label{App:Invariants.in.different.parametrizations}

Let us define three distinct parametrizations as in \cite{JKSV.Invariants,Jarv:2015kga,Vilson.Some,JKSV.Parametrizations}.
\begin{itemize}
 	\item The Jordan frame in Brans-Dicke-Bergmann-Wagoner parametrization (JF BDBW) 
 	\cite{Dicke:1962gz,BDBW} for the 
 	scalar field $\Psi$:
 	\begin{equation}
 	\label{JF.BDBW}
 	\mathcal{A} = \Psi \,, \qquad \mathcal{B} = \frac{\omega(\Psi)}{\Psi} \,, \qquad \mathcal{V} = \mathcal{V}(\Psi) \,, \qquad \alpha = 0 \,.
 	\end{equation}
 	
 	\item  The Jordan frame in Boisseau-Esposito-Far\`{e}se-Polarski-Starobinsky parametrization (JF BEPS) \cite{BEPS} for the scalar field $\phi$:
 	\begin{equation}
 	\label{JF.BEPS}
 	\mathcal{A} = F(\phi) \,, \qquad \mathcal{B} = 1 \,, \qquad \mathcal{V}  = \mathcal{V}(\phi) \,, \qquad \alpha = 0 \,.
 	\end{equation} 
 	
 	\item  The Einstein frame in canonical parametrization (EF canonical) \cite{Dicke:1962gz,BDBW} 
 	for the scalar field $\varphi$
 	\begin{equation}
 	\label{EF.canonical}
 	\mathcal{A} = 1 \,, \qquad \mathcal{B} = 2 \,, \qquad \mathcal{V} = \mathcal{V}(\varphi) \,, \qquad \alpha = \alpha(\varphi) \,. 
 	\end{equation}
\end{itemize}
The invariants (\ref{I.1.I.2})-(\ref{Invariant.metric2}) in different parametrizations are presented in table \ref{Table}.
 
\renewcommand{\arraystretch}{10}
\begin{table}[h]
 	\caption{\label{Table} Invariants in different parametrizations.}
 	\fontsize{7}{4}\selectfont
 	\begin{tabular}{@{}lllll}
		\toprule[2pt]
 		{\normalsize Invariant} & {\normalsize General parametrization} & {\normalsize JF BDBW} & {\normalsize JF BEPS} & {\normalsize EF can.} \\
 		\midrule[1pt]
 		%
 		%
 		$\mathcal{I}_1$ & $e^{2\alpha(\Phi)} \left( \mathcal{A}(\Phi) \right)^{-1}$ & $\Psi^{-1}$ & $\left( F(\phi) \right)^{-1}$ & $e^{2\alpha(\varphi)}$ \\
 		\hline
 		%
 		%
 		$\mathcal{I}_2$ & $\left( \mathcal{A}(\Phi) \right)^{-2}\mathcal{V}(\Phi)$ & $\Psi^{-2} \mathcal{V}(\Psi)$ & $\left( F(\phi) \right)^{-2} \mathcal{V}(\phi)$ & $\mathcal{V}(\varphi)$ \\
 		\hline
 		%
 		%
 		$\mathcal{I}_3$ & $\pm \mathop{\mathlarger{\mathlarger{\int}}} \left( \frac{2\mathcal{A}(\Phi)\mathcal{B}(\Phi) + 3 \left(\mathcal{A}^\prime(\Phi)\right)^2}{4\mathcal{A}(\Phi)^2} \right)^{\frac{1}{2}} \,\mathrm{d} \Phi$ & $\pm \mathop{\mathlarger{\mathlarger{\int}}} \left(\frac{2\omega(\Psi) + 3}{4\Psi^2}\right)^{\frac{1}{2}} \,\mathrm{d} \Psi$ & $\pm \mathop{\mathlarger{\mathlarger{\int}}} \left(\frac{2F(\phi) + 3 \left(F^\prime(\phi)\right)^2}{4F(\phi)^2}\right)^{\frac{1}{2}} \,\mathrm{d} \phi$ & $\pm\varphi + \mathit{const}$ \\
 		\hline
 		%
 		%
 		$\mathcal{I}_4\equiv \mathcal{I}_{1}^{-2}\mathcal{I}_2$ & $e^{-4\alpha(\Phi)}\mathcal{V}(\Phi)$ & $\mathcal{V}(\Psi)$ & $\mathcal{V}(\phi)$ & $e^{-4\alpha(\varphi)} \mathcal{V}(\varphi)$ \\
 		\hline
 		%
 		%
 		$\mathcal{I}_5\equiv \left(\frac{\mathcal{I}_1^\prime}{2\mathcal{I}_1\mathcal{I}_3^\prime}\right)^2$ & $\frac{\left(2\alpha^\prime(\Phi)\mathcal{A}(\Phi) - \mathcal{A}^\prime(\Phi)\right)^2}{2\mathcal{A}(\Phi)\mathcal{B}(\Phi) + 3\left(\mathcal{A}^\prime(\Phi)\right)^2}$ & $\frac{1}{2\omega(\Psi) + 3}$ & $\frac{\left(F^\prime(\phi)\right)^2}{2F(\phi) + 3\left(F^\prime(\phi)\right)^2}$ & $\left(\alpha^\prime(\varphi)\right)^2$ \\
 		\hline
 		%
 		%
 		$\bar{g}_{\mu\nu}$ & $e^{2\alpha(\Phi)}g_{\mu\nu}$ & $g_{\mu\nu}$ & $g_{\mu\nu}$ & $e^{2\alpha(\varphi)}g_{\mu\nu}$ \\
 		\hline 
 		%
 		%
 		$\hat{g}_{\mu\nu}$ & $\mathcal{A}(\Phi)g_{\mu\nu}$ & $\Psi g_{\mu\nu}$ & $F(\phi)g_{\mu\nu}$ & $g_{\mu\nu}$ \\
		\toprule[2pt]
	\end{tabular}
	\vspace{0.0cm}
\end{table}
\renewcommand{\arraystretch}{1} 
 		

\subsection{Slow-roll parameters in different parametrizations}\label{App:Slow.roll.parameters.in.different.parametrizations}


\subsubsection{General parametrization}

Let us write out the slow-roll parameters (\ref{kappa.JF}), (\ref{lambda.0.JF})-(\ref{lambda.1.JF}) in general parametrization \cite{Jarv:2015kga,Flanagan,JKSV.Parametrizations}.
%
\begin{subequations}
%
 \begin{eqnarray}
 \bar{\kappa}_0 &=&  \frac{ \mathcal{F} \dot{\Phi}^2 }{ (H + \dot{\alpha})^2 } \,, \qquad \qquad
 \bar{\kappa}_1 = -2\frac{ \dot{H} + \ddot{\alpha} }{ (H + \dot{\alpha})^2 } +
 \frac{ \dot{\mathcal{F}} }{ \mathcal{F} (H + \dot{\alpha}) } + 2\frac{ \ddot{\Phi} }{ \dot{\Phi} (H + \dot{\alpha}) }
 \,, \\
 \bar{\lambda}_0 &=& \frac{ 2\mathcal{A} \dot{\alpha} - \dot{\mathcal{A}} }{ 2 \mathcal{A} \left( H + \dot{\alpha} \right) } \,, \qquad \quad
 \bar{\lambda}_1 = -\frac{ \dot{H} + \ddot{\alpha} }{ (H + \dot{\alpha})^2 } +
 \frac{ 2 \mathcal{A}^2 \ddot{\alpha} - \mathcal{A} \ddot{\mathcal{A}} +  \dot{\mathcal{A}}^2 }{\mathcal{A} ( 2\mathcal{A}\dot{\alpha} - \dot{\mathcal{A}} ) (H + \dot{\alpha} ) }
 \,.
 \end{eqnarray}
%
\end{subequations}
Here $H$ denotes the Hubble parameter in general (arbitrary) parametrization and the dot represents the time derivative.


\subsubsection{Canonical Einstein frame}
In the canonical Einstein frame (\ref{EF.canonical}) the system (\ref{Friedmann1.EF})-(\ref{Friedmann3.EF}) or equivalently (up to an overall factor) the system (\ref{Friedmann1.JF})-(\ref{Friedmann3.JF}) reads
\begin{equation}
\label{Fr.EFcanonical} 
	H^2 = \frac{1}{3} \dot{\varphi}^2 + \frac{1}{ 3 \ell^2 } \mathcal{V} \,, \qquad \qquad 
	\dot{H} = -\dot{\varphi}^2 \,, \qquad \qquad
	\ddot{\varphi} = -3H\dot{\varphi} - \frac{1}{ 2 \ell^2 } \mathcal{V}^{\prime} \,.
\end{equation}
Here we use the notation where all quantities are defined in the Einstein frame. In that frame $\mathcal{F} = 1$. The slow-roll parameters (\ref{kappa.JF}), (\ref{lambda.0.JF})-(\ref{lambda.1.JF}) are given by
%
\begin{subequations}
%
\begin{eqnarray}
\bar{\kappa}_{0} &=& \frac{ \dot{\varphi}^2 }{ (H + \dot{\alpha})^2 } \,, \qquad \qquad
\bar{\kappa}_1 = -2\frac{ \dot{H} + \ddot{\alpha} }{ (H + \dot{\alpha})^2 } + 2\frac{ \ddot{\varphi} }{ \dot{\varphi} (H + \dot{\alpha}) } \,, \\
\bar{\lambda}_0 &=& \frac{ \dot{\alpha} }{ H + \dot{\alpha} } \,, \qquad \qquad \quad
\bar{\lambda}_1 = -\frac{ \dot{H} + \ddot{\alpha} }{ (H + \dot{\alpha})^2 } +	\frac{ \ddot{\alpha} }{ (H + \dot{\alpha}) \dot{\alpha} } = \frac{ \ddot{\alpha} H - \dot{H} \dot{\alpha} }{ (H + \dot{\alpha})^2 \dot{\alpha} } \,.
\end{eqnarray}
%
\end{subequations}
Now by imposing the conditions (\ref{slow.roll.JF}) we obtain from $|\bar{\lambda}_0| \ll 1$ that $|\dot{\alpha}| \ll H$. Using the latter in the expression for $\bar{\kappa}_0$ we deduce that $3\dot{\varphi}^2 \ll 3 H^2 \approx  \ell^{-2}\mathcal{V}$, which is the familiar slow-roll condition. This condition guarantees also that $|\dot{H}| \ll H^2$ holds. From $|\bar{\lambda}_1| \ll 1$ we obtain $|\ddot{\alpha}| \ll |\dot{\alpha}|H$ and $|\ddot{\alpha}| \ll H^2$. Using the latter in the expression for $\bar{\kappa}_1$ we deduce that $|\ddot{\varphi}| \ll H |\dot{\varphi}|$. Thus we have obtained the familiar conditions and approximation
\begin{equation}
\label{sr.JF.EFcanonical1}
H^2 \approx \frac{1}{3 \ell^2}\mathcal{V} \,, \qquad \qquad
3H \dot{\varphi} \approx - \frac{1}{ 2 \ell^2 } \mathcal{V}^{\,\prime}
\end{equation}
as in  the standard slow-roll.


\subsubsection{Jordan frame BEPS}

In Jordan frame BEPS parametrization (\ref{JF.BEPS}) the system (\ref{Friedmann1.EF})-(\ref{Friedmann3.EF}) (or equivalently (\ref{Friedmann1.JF})-(\ref{Friedmann3.JF})) reads
\begin{subequations}
%
\begin{eqnarray} 
H^2 &=& -H \frac{ \dot{F} }{ F } + \frac{ 1 }{ 6F } \dot{\phi}^2 + \frac{1}{ 3 \ell^2 } \frac{ \mathcal{V} }{ F }  \,, \\
\label{Friedmann.2.equation.BEPS}
\dot{H} &=& H \frac{ \dot{F} }{ 2 F } - \frac{ 1 }{ 2F } \dot{\phi}^2  -
\frac{\ddot{F}}{2F} \,, \\
\label{scalar.field.equation.BEPS}
\ddot{\phi} + \frac{ F^{\prime} \left( 1 + 3 F^{\prime\prime} \right) }{ 2F + 3 \left( F^{\prime} \right)^2 } \dot{\phi}^2 &=& - 3 H \dot{\phi} - \frac{ 2 F^3 }{ \ell^2 \left( 2F + 3 \left( F^{\prime} \right)^2 \right) } \left( \frac{\mathcal{V}}{F^2} \right)^{\prime}
\,.
\end{eqnarray}
%
\end{subequations}
Here we use the notation where all quantities are defined explicitly in the Jordan frame. The slow-roll parameters  (\ref{kappa.JF}), (\ref{lambda.0.JF})-(\ref{lambda.1.JF}) are given by
%
\begin{subequations}
	%
	\begin{eqnarray}
	\bar{\kappa}_0 &=&
	\frac{ \dot{\phi}^2 }{ H^2 } \frac{ 2F + 3(F^{\prime})^2 }{ 4F^2 } \,, \qquad
	\bar{\kappa}_1 = -2 \frac{ \dot{H} }{ H^2 } - 2 \frac{ \dot{F} }{ F H } + 2 \frac{ \ddot{\phi} }{ \dot{\phi} H } + 2 \frac{ F^{\prime} \left( 1 + 3 F^{\prime\prime} \right) \dot{\phi} }{ \left( 2F + 3 \left( F^{\prime} \right)^2 \right) H } \,,
	\\
	\bar{\lambda}_0 &=& -\frac{\dot{F}}{2 F H} \,, \qquad \qquad \qquad
	\bar{\lambda}_1 = -\frac{ \dot{H} }{ H^2 } + \frac{ \ddot{F} }{ \dot{F} H } - \frac{ \dot{F} }{ FH } \,.
	\end{eqnarray}
	%
\end{subequations}
Now by imposing the conditions (\ref{slow.roll.JF}) we obtain from $|\bar{\lambda}_0| \ll 1$ that
$|\dot{F}| \ll FH$. Using this in the expression for $\bar{\kappa}_0$ we deduce that $|\dot{\phi}^2|\ll FH^2$.  From $|\bar{\lambda}_1| \ll 1$ it follows that the difference $\ddot{F}/\dot{F}H - \dot{H}/H^2$ must be small which due to underlying assumptions is consistent with (\ref{Friedmann.2.equation.BEPS}) only if $|\ddot{F}| \ll |\dot{F}|H$ and $|\dot{H}| \ll H^2$. Hence, $|\bar{\kappa}_1| \ll 1$ states that the l.h.s.\ of (\ref{scalar.field.equation.BEPS}) is small. If we further impose $\mathcal{O}(\bar{\lambda}_0) = \mathcal{O}(\bar{\kappa}_0)$ then $|F^{\prime}| \ll \sqrt{F}$ and we obtain the approximation
\begin{equation}
\label{sr.JF.Beps1}
H^2 \approx \frac{1}{ 3 \ell^2 } \frac{\mathcal{V}}{F} \,, \qquad \qquad 
3H\dot{\phi} \approx - \frac{ F^2 }{ \ell^2 } \left( \frac{\mathcal{V}}{F^2} \right)^{\prime} \,.
\end{equation}
From the same assumption, by comparing the expressions for $\bar{\lambda}_{1}$ and $\bar{\kappa}_{1}$, we deduce that $|\ddot{\phi}| \ll |\dot{\phi}|H$. This is consistent with the calculations performed in \cite{Chiba:2008ia}.





\begin{thebibliography}{99}

%
%
%


\bibitem{Guth:1980zm}
  Guth A H 1981 Inflationary universe: a possible solution to the horizon and flatness problems {\it Phys.\ Rev.} D \href{http://dx.doi.org/10.1103/PhysRevD.23.347}{{\bf 23} 347--56}
  %
  {\hypersetup{urlcolor=cyan}
  [\href{http://inspirehep.net/record/154280}{inSPIRE}]}
     
  
\bibitem{Linde:1981mu}
  Linde A D 1982 A new inflationary universe scenario: a possible solution of the horizon, flatness, homogeneity, isotropy and primordial monopole problems {\it Phys.\ Lett.} B \href{http://dx.doi.org/10.1016/0370-2693(82)91219-9}{{\bf 108} 389--93}
  %
  {\hypersetup{urlcolor=cyan}
  [\href{http://inspirehep.net/record/168781}{inSPIRE}]}
  \\
  %
  %
  %
  Linde A D 1983 Chaotic inflation {\it Phys.\ Lett.} B \href{http://dx.doi.org/10.1016/0370-2693(83)90837-7}{{\bf 129} 177--81}
  %
  {\hypersetup{urlcolor=cyan}
  [\href{http://inspirehep.net/record/196244}{inSPIRE}]}
  
  
\bibitem{Linde}
  Linde A D  1990 {\it Particle Physics and Inflationary Cosmology} ({\it Contemp.\ Concepts Phys.}~5) (Chur: Harwood Academic Publishers)
  %
  {\hypersetup{urlcolor=violet}
  (arXiv:\href{https://arxiv.org/abs/hep-th/0503203}{hep-th/0503203})}
  %
  {\hypersetup{urlcolor=cyan}
  [\href{https://inspirehep.net/record/680653}{inSPIRE}]}

  
\bibitem{Steinhardt:1984}
  Steinhardt P J and Turner M S 1984 Prescription for successful new inflation {\it Phys.\ Rev.} D \href{http://dx.doi.org/10.1103/PhysRevD.29.2162}{{\bf 29} 2162--71}
  %
  {\hypersetup{urlcolor=cyan}
  [\href{http://inspirehep.net/record/201201}{inSPIRE}]}
  

\bibitem{Mukhanov:2005sc}
  Mukhanov V 2005 {\it Physical Foundations of Cosmology} (New York: Cambridge University Press)
  %
  {\hypersetup{urlcolor=cyan}
  [\href{http://inspirehep.net/record/706151}{inSPIRE}]}
  
 
\bibitem{Lyth:2009zz}
  Lyth D H and Riotto A 1999 Particle physics models of inflation and the cosmological density perturbation {\it Phys.\ Rep.} \href{http://dx.doi.org/10.1016/S0370-1573(98)00128-8}{{\bf 314} 1--146}
  %
  {\hypersetup{urlcolor=violet}
  	(arXiv:\href{https://arxiv.org/abs/hep-ph/9807278}{hep-ph/9807278})}
  %
  {\hypersetup{urlcolor=cyan}
  	[\href{https://inspirehep.net/record/472934}{inSPIRE}]}
  \\
  %
  %
  %
  Lyth D H and Liddle A R 2009 {\it The Primordial Density Perturbation: Cosmology, Inflation and the Origin of Structure} (New York: Cambridge University Press)
  %
  {\hypersetup{urlcolor=cyan}
  [\href{http://inspirehep.net/record/853992}{inSPIRE}]}


\bibitem{Liddle:1994dx} 
  Liddle A R, Parsons P and Barrow J D 1994 Formalizing the slow-roll approximation in inflation {\it Phys.\ Rev.} D \href{http://dx.doi.org/10.1103/PhysRevD.50.7222}{{\bf 50} 7222--32}
  %
  {\hypersetup{urlcolor=violet}
  (arXiv:\href{https://arxiv.org/abs/astro-ph/9408015}{astro-ph/9408015})}
  %
  {\hypersetup{urlcolor=cyan}
  [\href{https://inspirehep.net/record/375405}{inSPIRE}]}
  
  
\bibitem{Ijjas:2014nta} 
  Ijjas A, Steinhardt P J and Loeb A 2014 Inflationary schism {\it Phys.\ Lett.} B \href{http://dx.doi.org/10.1016/j.physletb.2014.07.012}{{\bf 736} 142--6}
  %
  {\hypersetup{urlcolor=violet}
  (arXiv:\href{https://arxiv.org/abs/1402.6980}{1402.6980})} 
  %
  {\hypersetup{urlcolor=cyan}
  [\href{https://inspirehep.net/record/1282879}{inSPIRE}]}


\bibitem{Ade:2015lrj} 
  Ade P A R {\it et al} (Planck Collaboration) 2016 Planck 2015 results. XX. Constraints on inflation {\it A\&A} \href{http://dx.doi.org/10.1051/0004-6361/201525898}{{\bf 594} A20}
  %
  {\hypersetup{urlcolor=violet}
  (arXiv:\href{https://arxiv.org/abs/1502.02114}{1502.02114})} 
  %
  {\hypersetup{urlcolor=cyan}
  [\href{https://inspirehep.net/record/1343460}{inSPIRE}]}


\bibitem{Starobinsky:1980te} 
  Starobinsky A A 1980 A new type of isotropic cosmological models without singularity {\it Phys.\ Lett.} B \href{http://dx.doi.org/10.1016/0370-2693(80)90670-X}{{\bf 91} 99--102}
  %
  {\hypersetup{urlcolor=cyan}
  [\href{http://inspirehep.net/record/157549}{inSPIRE}]}

  
\bibitem{Bezrukov:2007ep} 
  Bezrukov F and Shaposhnikov M 2008 The Standard Model Higgs boson as the inflaton {\it Phys.\ Lett.} B \href{http://dx.doi.org/10.1016/j.physletb.2007.11.072}{{\bf 659} 703–6}
  %
  {\hypersetup{urlcolor=violet}
  (arXiv:\href{https://arxiv.org/abs/0710.3755}{0710.3755})} 
  %
  {\hypersetup{urlcolor=cyan}
  [\href{https://inspirehep.net/record/764869}{inSPIRE}]}
  \\
  %
  %
  %
  Steinwachs C F 2014 {\it Non-minimal Higgs Inflation and Frame Dependence in Cosmology} ({\it Springer Theses}) (New York: Springer) 
  %
  DOI:\href{http://dx.doi.org/10.1007/978-3-319-01842-3}{10.1007/978-3-319-01842-3}
  %
  {\hypersetup{urlcolor=cyan}
  [\href{http://inspirehep.net/record/1251588}{inSPIRE}]}


\bibitem{Kallosh:2013hoa} 
  Kallosh R and Linde A 2013 Universality class in conformal inflation {\it J.\ Cosmol.\ Astropart.\ Phys.} \href{http://dx.doi.org/10.1088/1475-7516/2013/07/002}{JCAP07(2013)002}
  %
  {\hypersetup{urlcolor=violet}
  (arXiv:\href{https://arxiv.org/abs/1306.5220}{1306.5220})} 
  %
  {\hypersetup{urlcolor=cyan}
  [\href{https://inspirehep.net/record/1239479}{inSPIRE}]}
  \\
  %
  %
  %
  Kallosh R and Linde A 2013 Non-minimal inflationary attractors {\it J.\ Cosmol.\ Astropart.\ Phys.} \href{http://dx.doi.org/10.1088/1475-7516/2013/10/033}{JCAP10(2013)033}
  %
  {\hypersetup{urlcolor=violet}
  (arXiv:\href{https://arxiv.org/abs/1307.7938}{1307.7938})} 
  %
  {\hypersetup{urlcolor=cyan}
  [\href{https://inspirehep.net/record/1245204}{inSPIRE}]}
  \\
  %
  %
  %
  Carrasco J J M, Kallosh R and Linde A 2015 Cosmological attractors and initial conditions for inflation {\it Phys.\ Rev.} D \href{http://dx.doi.org/10.1103/PhysRevD.92.063519}{{\bf 92} 063519}
  %
  {\hypersetup{urlcolor=violet}
  (arXiv:\href{https://arxiv.org/abs/1506.00936}{1506.00936})} 
  %
  {\hypersetup{urlcolor=cyan}
  [\href{https://inspirehep.net/record/1374195}{inSPIRE}]}


\bibitem{Barrow}
  Barrow J D 1995 Slow-roll inflation in scalar-tensor theories {\it Phys.\ Rev.} D \href{http://dx.doi.org/10.1103/PhysRevD.51.2729}{{\bf 51} 2729--32}
  %
  {\hypersetup{urlcolor=cyan}
	[\href{https://inspirehep.net/record/406285}{inSPIRE}]}
  \\
  %
  %
  %
  Garc\'{\i}a-Bellido J and Wands D 1995 Constraints from inflation on scalar-tensor gravity theories {\it Phys.\ Rev.} D \href{http://dx.doi.org/10.1103/PhysRevD.52.6739}{{\bf 52} 6739--49} 
  %
  {\hypersetup{urlcolor=violet}
  	(\href{https://arxiv.org/abs/gr-qc/9506050}{gr-qc/9506050})}
  %
  {\hypersetup{urlcolor=cyan}
  	[\href{http://inspirehep.net/record/396571}{inSPIRE}]}
  

\bibitem{Faraoni:1998qx}
  Faraoni V, Gunzig E and Nardone P 1999 Conformal transformations in classical gravitational theories and in cosmology {\it Fundam.\ Cosm.\ Phys.} {\bf 20} 121
  %
  {\hypersetup{urlcolor=violet}
  (\href{https://arxiv.org/abs/gr-qc/9811047}{gr-qc/9811047})}
  %
  {\hypersetup{urlcolor=cyan}
  [\href{https://inspirehep.net/record/479326}{inSPIRE}]}


\bibitem{Capozziello:2010zz}
  Faraoni V and Capozziello S 2011 {\it Beyond Einstein Gravity : A Survey of Gravitational Theories for Cosmology and Astrophysics (Fundamental Theories of Physics Volume 170)}  (Dordrecht: Springer Netherlands) DOI:\href{http://dx.doi.org/10.1007/978-94-007-0165-6}{10.1007/978-94-007-0165-6} 
  %
  {\hypersetup{urlcolor=cyan}
  [\href{http://inspirehep.net/record/1107700}{inSPIRE}]}

  
\bibitem{Chiba:2014sva} 
  Chiba T and Kohri K 2015 Consistency relations for large-field inflation: non-minimal coupling {\it Prog.\ Theor.\ Exp.\ Phys.}
  \href{http://dx.doi.org/10.1093/ptep/ptv007}{(2015) 023E01} 
  %
  {\hypersetup{urlcolor=violet}
  	(arXiv:\href{https://arxiv.org/abs/1411.7104}{1411.7104})} 
  %
  {\hypersetup{urlcolor=cyan}
  	[\href{https://inspirehep.net/record/1330256}{inSPIRE}]}
  

\bibitem{Clifton:2011jh}
   Nojiri S and Odintsov S D 2011 Unified cosmic history in modified gravity: from $F(R)$ theory to Lorentz non-invariant models {\it Phys.\ Rep.} \href{http://dx.doi.org/10.1016/j.physrep.2011.04.001}{{\bf 505} 59--144}
   %
   {\hypersetup{urlcolor=violet}
   	(arXiv:\href{https://arxiv.org/abs/1011.0544}{1011.0544})} 
   %
   {\hypersetup{urlcolor=cyan}
   	[\href{https://inspirehep.net/record/875254}{inSPIRE}]}
   \\
   %
   %
   %
  Clifton T, Ferreira P G, Padilla A and Skordis C 2012 Modified gravity and cosmology {\it Phys.\ Rep.} \href{http://dx.doi.org/10.1016/j.physrep.2012.01.001}{{\bf 513} 1--189}
  %
  {\hypersetup{urlcolor=violet}
  	(arXiv:\href{https://arxiv.org/abs/1106.2476}{1106.2476})} 
  %
  {\hypersetup{urlcolor=cyan}
  	[\href{https://inspirehep.net/record/913415}{inSPIRE}]}
  \\
  %
  %
  %
  Bull P {\it et al} 2016 Beyond $\Lambda$CDM: problems, solutions, and the road ahead {\it Phys.\ Dark Univ.} \href{http://dx.doi.org/10.1016/j.dark.2016.02.001}{{\bf 12} 56--99}
  %
  {\hypersetup{urlcolor=violet}
  	(arXiv:\href{https://arxiv.org/abs/1512.05356}{1512.05356})} 
  %
  {\hypersetup{urlcolor=cyan}
  	[\href{https://inspirehep.net/record/1410025}{inSPIRE}]}


\bibitem{Faraoni.Sotiriou}
  O'Hanlon J 1972 Intermediate-range gravity: a generally covariant model {\it Phys.\ Rev.\ Lett.} \href{http://dx.doi.org/10.1103/PhysRevLett.29.137}{{\bf 29} 137--8}
  %
  {\hypersetup{urlcolor=cyan}
	[\href{https://inspirehep.net/record/75459}{inSPIRE}]} 
  \\
  %
  %
  %
  Chiba T 2003 $1/R$ gravity and scalar-tensor gravity {\it Phys.\ Lett.} B \href{http://dx.doi.org/10.1016/j.physletb.2003.09.033}{{\bf 575} 1--3}
  %
  {\hypersetup{urlcolor=violet}
  	(arXiv:\href{https://arxiv.org/abs/astro-ph/0307338}{astro-ph/0307338})}
  %
  {\hypersetup{urlcolor=cyan}
  	[\href{https://inspirehep.net/record/623724}{inSPIRE}]}
  \\
  %
  %
  %
  Capozziello S, Nojiri S, Odintsov S D and Troisi A 2006 Cosmological viability of $f(R)$-gravity as an ideal fluid and its compatibility with a matter dominated phase {\it Phys.\ Lett.} B \href{http://dx.doi.org/10.1016/j.physletb.2006.06.034}{{\bf 639} 135--43}
  %
  {\hypersetup{urlcolor=violet}   	(arXiv:\href{https://arxiv.org/abs/astro-ph/0604431}{astro-ph/0604431})}
  %
  {\hypersetup{urlcolor=cyan}
  	[\href{https://inspirehep.net/record/714877}{inSPIRE}]}  
  \\
  %
  %
  %
  Sotiriou T P and Faraoni V 2010 $f(R)$ theories of gravity {\it Rev.\ Mod.\ Phys.} \href{http://dx.doi.org/10.1103/RevModPhys.82.451}{{\bf 82} 451--97}
  %
  {\hypersetup{urlcolor=violet}
  	(arXiv:\href{https://arxiv.org/abs/0805.1726}{0805.1726})} 
  %
  {\hypersetup{urlcolor=cyan}
  	[\href{https://inspirehep.net/record/785578}{inSPIRE}]}


\bibitem{Shaposhnikov:2008xi} 
  Shaposhnikov M and Zenh\"ausern D 2009 Quantum scale invariance, cosmological constant and hierarchy problem {\it Phys.\ Lett.} B \href{http://dx.doi.org/10.1016/j.physletb.2008.11.041}{{\bf 671} 162--6}
  %
  {\hypersetup{urlcolor=violet}
  	(arXiv:\href{https://arxiv.org/abs/0809.3406}{0809.3406})} 
  %
  {\hypersetup{urlcolor=cyan}
  	[\href{https://inspirehep.net/record/797091}{inSPIRE}]}
  \\
  %
  %
  %
  Kannike K, H\"utsi G, Pizza L, Racioppi A, Raidal M, Salvio A and Strumia A 2015 Dynamically induced Planck scale and inflation {\it J.\ High Energy Phys.} \href{http://dx.doi.org/10.1007/JHEP05(2015)065}{JHEP05(2015)065}
  %
  {\hypersetup{urlcolor=violet}
  	(arXiv:\href{https://arxiv.org/abs/1502.01334}{1502.01334})} 
  %
  {\hypersetup{urlcolor=cyan}
  	[\href{https://inspirehep.net/record/1342926}{inSPIRE}]}
  \\
  %
  %
  %
  Ferreira P G, Hill C T and Ross G G 2016 Scale-independent inflation and hierarchy generation
  %
  {\hypersetup{urlcolor=violet}
  	arXiv:\href{https://arxiv.org/abs/1603.05983}{1603.05983}} 
  %
  {\hypersetup{urlcolor=cyan}
  	[\href{https://inspirehep.net/record/1430849}{inSPIRE}]}


\bibitem{sasaki:nonlocal}
  De Felice A and Sasaki M 2015 Ghosts in classes of non-local gravity {\it Phys.\ Lett.} B \href{http://dx.doi.org/10.1016/j.physletb.2015.02.045}{{\bf 743} 189--97}
  %
  {\hypersetup{urlcolor=violet}
  	(arXiv:\href{https://arxiv.org/abs/1412.1575}{1412.1575})} 
  %
  {\hypersetup{urlcolor=cyan}
  	[\href{https://inspirehep.net/record/1332718}{inSPIRE}]}


\bibitem{Dicke:1962gz}
  Dicke R H 1962 Mach's principle and invariance under transformation of units {\it Phys.\ Rev.} \href{http://dx.doi.org/10.1103/PhysRev.125.2163}{{\bf 125} 2163--7}
  %
  {\hypersetup{urlcolor=cyan}
  [\href{http://inspirehep.net/record/2035}{inSPIRE}]}


\bibitem{physical.equivalence}
  Faraoni V and Nadeau S 2007 (Pseudo)issue of the conformal frame revisited {\it Phys.\ Rev.} D
  \href{http://dx.doi.org/10.1103/PhysRevD.75.023501}{{\bf 75} 023501}
  %
  {\hypersetup{urlcolor=violet}
  	(arXiv:\href{https://arxiv.org/abs/gr-qc/0612075}{gr-qc/0612075})}
  %
  {\hypersetup{urlcolor=cyan}
  	[\href{https://inspirehep.net/record/734354}{inSPIRE}]}
  \\
  %
  %
  %
  Briscese F, Elizalde E, Nojiri S and Odintsov S D 2007 Phantom scalar dark energy as modified gravity: understanding the origin of the Big Rip singularity {\it Phys.\ Lett.} B \href{http://dx.doi.org/10.1016/j.physletb.2007.01.013}{{\bf 646} 105--11} 
  %
  {\hypersetup{urlcolor=violet}
  	(arXiv:\href{https://arxiv.org/abs/hep-th/0612220}{hep-th/0612220})}
  %
  {\hypersetup{urlcolor=cyan}
  	[\href{https://inspirehep.net/record/735208}{inSPIRE}]}  
  \\
  %
  %
  %
  Bhadra A, Sarkar K, Datta D P and Nandi K K 2007 Brans-Dicke theory: Jordan versus Einstein frame {\it Mod.\ Phys.\ Lett.} A \href{http://dx.doi.org/10.1142/S021773230702261X}{{\bf 22} 367--76}
  %
  {\hypersetup{urlcolor=violet}
  	(arXiv:\href{https://arxiv.org/abs/gr-qc/0605109}{gr-qc/0605109})}
  %
  {\hypersetup{urlcolor=cyan}
  	[\href{https://inspirehep.net/record/717368}{inSPIRE}]}
  \\
  %
  %
  %
  Nozari K and Sadatian S D 2009 Comparison of frames: Jordan versus Einstein frame for a non-minimal dark energy model {\it Mod.\ Phys.\ Lett.} A \href{http://dx.doi.org/10.1142/S0217732309031053}{{\bf 24} 3143--55}
  %
  {\hypersetup{urlcolor=violet}
  (arXiv:\href{https://arxiv.org/abs/0905.0241}{0905.0241})} 
  %
  {\hypersetup{urlcolor=cyan}
  [\href{https://inspirehep.net/record/819320}{inSPIRE}]}
  \\
  %
  %
  %
  Capozziello S, Martin-Moruno P and Rubano C 2010 Physical non-equivalence of the Jordan and Einstein frames {\it Phys.\ Lett.} B \href{http://dx.doi.org/10.1016/j.physletb.2010.04.058}{{\bf 689} 117--21}
  %
  {\hypersetup{urlcolor=violet}
  (arXiv:\href{https://arxiv.org/abs/1003.5394}{1003.5394})} 
  %
  {\hypersetup{urlcolor=cyan}
  [\href{https://inspirehep.net/record/850278}{inSPIRE}]}
  \\
  %
  %
  %
  Corda C 2011 Gravitational wave astronomy: the definitive test for the ``Einstein frame versus Jordan frame'' controversy {\it } {\it Astropart.\ Phys.} \href{http://dx.doi.org/10.1016/j.astropartphys.2010.10.006}{{\bf 34} 412--9}
  %
  {\hypersetup{urlcolor=violet}
  (arXiv:\href{https://arxiv.org/abs/1010.2086}{1010.2086})} 
  %
  {\hypersetup{urlcolor=cyan}
  [\href{https://inspirehep.net/record/872583}{inSPIRE}]}
  \\
  %
  %
  %
  Stabile A, Stabile An and Capozziello S 2013 Conformal transformations and weak field limit of scalar-tensor gravity {\it Phys.\ Rev.} D \href{http://dx.doi.org/10.1103/PhysRevD.88.124011}{{\bf 88} 124011}
  %
  {\hypersetup{urlcolor=violet}
  (arXiv:\href{https://arxiv.org/abs/1310.7097}{1310.7097})} 
  %
  {\hypersetup{urlcolor=cyan}
  [\href{https://inspirehep.net/record/1262278}{inSPIRE}]}
  \\
  %
  %
  %
  Obukhov Y N and Puetzfeld D 2014 Equations of motion in scalar-tensor theories of gravity: a covariant multipolar approach {\it Phys.\ Rev.} D \href{http://dx.doi.org/10.1103/PhysRevD.90.104041}{{\bf 90} 104041}
  %
  {\hypersetup{urlcolor=violet}
  (arXiv:\href{https://arxiv.org/abs/1404.6977}{1404.6977})} 
  %
  {\hypersetup{urlcolor=cyan}
  [\href{https://inspirehep.net/record/1292784}{inSPIRE}]}
  \\
  %
  %
  %
  Bahamonde S, Odintsov S D, Oikonomou V K and Wright M 2016 Correspondence of $F(R)$ gravity singularities in Jordan and Einstein frames {\it Annals Phys.} \href{http://dx.doi.org/10.1016/j.aop.2016.06.020}{{\bf 373} 96--114}
  %
  {\hypersetup{urlcolor=violet}
  	(arXiv:\href{https://arxiv.org/abs/1603.05113}{1603.05113})} 
  %
  {\hypersetup{urlcolor=cyan}
  	[\href{https://inspirehep.net/record/1428643}{inSPIRE}]}  
  

\bibitem{Chiba:2013mha} 
  Chiba T and Yamaguchi M 2013 Conformal-frame (in)dependence of cosmological observations in scalar-tensor theory {\it J.\ Cosmol.\ Astropart.\ Phys.} \href{http://dx.doi.org/10.1088/1475-7516/2013/10/040}{JCAP10(2013)040}
  %
  {\hypersetup{urlcolor=violet}
  	(arXiv:\href{https://arxiv.org/abs/1308.1142}{1308.1142})} 
  %
  {\hypersetup{urlcolor=cyan}
  	[\href{https://inspirehep.net/record/1246738}{inSPIRE}]}
  
  
\bibitem{Postma:2014vaa}
  Postma M and Volponi M 2014 Equivalence of the Einstein and Jordan frames {\it Phys.\ Rev.} D \href{http://dx.doi.org/10.1103/PhysRevD.90.103516}{{\bf 90} 103516}
  %
  {\hypersetup{urlcolor=violet}
  	(arXiv:\href{https://arxiv.org/abs/1407.6874}{1407.6874})} 
  %
  {\hypersetup{urlcolor=cyan}
  	[\href{https://inspirehep.net/record/1308081}{inSPIRE}]}


\bibitem{Higgs.Inflation.and.Naturalness}
  Lerner R N and McDonald J 2010 Higgs inflation and naturalness {\it J.\ Cosmol.\ Astropart.\ Phys.} \href{http://dx.doi.org/10.1088/1475-7516/2010/04/015}{JCAP04(2010)015}
  %
  {\hypersetup{urlcolor=violet}
  (arXiv:\href{https://arxiv.org/abs/0912.5463}{0912.5463})} 
  %
  {\hypersetup{urlcolor=cyan}
  [\href{https://inspirehep.net/record/841416}{inSPIRE}]}
  

\bibitem{vandeBruck:2015gjd}
  van de Bruck C and Longden C 2016 Higgs inflation with a Gauss-Bonnet term in the Jordan frame {\it Phys.\ Rev.} D \href{http://dx.doi.org/10.1103/PhysRevD.93.063519}{{\bf 93} 063519}
  %
  {\hypersetup{urlcolor=violet}
  (arXiv:\href{https://arxiv.org/abs/1512.04768}{1512.04768})} 
  %
  {\hypersetup{urlcolor=cyan}
  [\href{https://inspirehep.net/record/1409713}{inSPIRE}]}
  
  
\bibitem{ArmendarizPicon:2015dma}
  Armendariz-Picon C 2015 On the scale of inflation
  %
  {\hypersetup{urlcolor=violet}
  arXiv:\href{https://arxiv.org/abs/1510.07956}{1510.07956}}
  %
  {\hypersetup{urlcolor=cyan}
  [\href{https://inspirehep.net/record/1400987}{inSPIRE}]}
  

\bibitem{Morris:2001ad}
  Morris J R 2001 Generalized slow-roll conditions and the possibility of intermediate scale inflation in scalar-tensor theory {\it Class.\ Quantum Grav.} \href{http://dx.doi.org/10.1088/0264-9381/18/15/311}{{\bf 18} 2977--88}
  %
  {\hypersetup{urlcolor=violet}
  (arXiv:\href{https://arxiv.org/abs/gr-qc/0106022}{gr-qc/0106022})}
  %
  {\hypersetup{urlcolor=cyan}
  [\href{https://inspirehep.net/record/557814}{inSPIRE}]}


\bibitem{Chiba:2008ia} 
  Chiba T and Yamaguchi M 2008 Extended slow-roll conditions and rapid-roll conditions {\it J.\ Cosmol.\ Astropart.\ Phys.} \href{http://dx.doi.org/10.1088/1475-7516/2008/10/021}{JCAP10(2008)021}
  %
  {\hypersetup{urlcolor=violet}
  (arXiv:\href{https://arxiv.org/abs/0807.4965}{0807.4965})} 
  %
  {\hypersetup{urlcolor=cyan}
  [\href{https://inspirehep.net/record/791874}{inSPIRE}]}


\bibitem{Torres:1996fr} 
  Torres D F 1997 Slow roll inflation in nonminimally coupled theories: hyperextended gravity approach {\it Phys.\ Lett.} A \href{http://dx.doi.org/10.1016/S0375-9601(96)00835-3}{{\bf 225} 13--7}
  %
  {\hypersetup{urlcolor=violet}
  (arXiv:\href{https://arxiv.org/abs/gr-qc/9610021}{gr-qc/9610021})}
  %
  {\hypersetup{urlcolor=cyan}
  [\href{https://inspirehep.net/record/424500}{inSPIRE}]}

  
\bibitem{Kaiser:1994vs} 
  Kaiser D I 1995 Primordial spectral indices from generalized Einstein theories {\it Phys.\ Rev.} D \href{http://dx.doi.org/10.1103/PhysRevD.52.4295}{{\bf 52} 4295--306}
  %
  {\hypersetup{urlcolor=violet}
  (arXiv:\href{https://arxiv.org/abs/astro-ph/9408044}{astro-ph/9408044})}
  %
  {\hypersetup{urlcolor=cyan}
  [\href{https://inspirehep.net/record/375629}{inSPIRE}]}


\bibitem{Kaiser:1995nv} 
  Kaiser D I 1995 Frame-independent calculation of spectral indices from inflation 
  %
  {\hypersetup{urlcolor=violet}
  arXiv:\href{https://arxiv.org/abs/astro-ph/9507048}{astro-ph/9507048}}
  %
  {\hypersetup{urlcolor=cyan}
  [\href{https://inspirehep.net/record/397184}{inSPIRE}]}

  
\bibitem{Noh:2001ia} 
  Noh H and Hwang J-C 2001 Inflationary spectra in generalized gravity: unified forms {\it Phys.\ Lett.} B \href{http://dx.doi.org/10.1016/S0370-2693(01)00875-9}{{\bf 515} 231--7}
  %
  {\hypersetup{urlcolor=violet}
  (arXiv:\href{https://arxiv.org/abs/astro-ph/0107069}{astro-ph/0107069})}
  %
  {\hypersetup{urlcolor=cyan}
  [\href{https://inspirehep.net/record/559516}{inSPIRE}]}

  
\bibitem{JKSV.Invariants} 
  J\"arv L, Kuusk P, Saal M and Vilson O 2015 Invariant quantities in the scalar-tensor theories of gravitation {\it Phys.\ Rev.} D \href{http://dx.doi.org/10.1103/PhysRevD.91.024041}{{\bf 91} 024041}
  %
  {\hypersetup{urlcolor=violet}
  (arXiv:\href{https://arxiv.org/abs/1411.1947}{1411.1947})} 
  %
  {\hypersetup{urlcolor=cyan}
  [\href{https://inspirehep.net/record/1326631}{inSPIRE}]}


\bibitem{Jarv:2015kga} 
  J\"arv L, Kuusk P, Saal M and Vilson O 2015 Transformation properties and general relativity regime in scalar–tensor theories {\it Class.\ Quantum Grav.} \href{http://dx.doi.org/10.1088/0264-9381/32/23/235013}{{\bf 32} 235013}
  %
  {\hypersetup{urlcolor=violet}
  (arXiv:\href{https://arxiv.org/abs/1504.02686}{1504.02686})} 
  %
  {\hypersetup{urlcolor=cyan}
  [\href{https://inspirehep.net/record/1358926}{inSPIRE}]}

  
\bibitem{Wetterich:2015ccd} 
  Wetterich C 2016 Primordial cosmic fluctuations for variable gravity {\it J.\ Cosmol.\ Astropart.\ Phys.} \href{http://dx.doi.org/10.1088/1475-7516/2016/05/041}{JCAP05(2016)041}
  %
  {\hypersetup{urlcolor=violet}
  (arXiv:\href{https://arxiv.org/abs/1511.03530}{1511.03530})} 
  %
  {\hypersetup{urlcolor=cyan}
  [\href{https://inspirehep.net/record/1403964}{inSPIRE}]}


\bibitem{Nozari:2010uu} 
  Nozari K and Shafizadeh S 2010 Non-minimal inflation revisited {\it Phys.\ Scr.} \href{http://dx.doi.org/10.1088/0031-8949/82/01/015901}{{\bf 82} 015901}
  %
  {\hypersetup{urlcolor=violet}
  (arXiv:\href{https://arxiv.org/abs/1006.1027}{1006.1027})} 
  %
  {\hypersetup{urlcolor=cyan}
  [\href{https://inspirehep.net/record/857285}{inSPIRE}]}

   
\bibitem{Flanagan} 
  Flanagan \'E \'E 2004 The conformal frame freedom in theories of gravitation {\it Class.\ Quantum Grav.}
  \href{http://dx.doi.org/10.1088/0264-9381/21/15/N02}{{\bf 21} 3817–29}
  %
  {\hypersetup{urlcolor=violet}
  (arXiv:\href{https://arxiv.org/abs/gr-qc/0403063}{gr-qc/0403063})}
  %
  {\hypersetup{urlcolor=cyan}
  [\href{https://inspirehep.net/record/646324}{inSPIRE}]}
  

\bibitem{Burns.et.al}
  Burns D, Karamitsos S and Pilaftsis A 2016 Frame-covariant formulation of inflation in scalar-curvature theories {\it Nucl.\ Phys.} B \href{http://dx.doi.org/10.1016/j.nuclphysb.2016.04.036}{{\bf 907} 785--819}
  %
  {\hypersetup{urlcolor=violet}
	(arXiv:\href{https://arxiv.org/abs/1603.03730}{1603.03730})} 
  %
  {\hypersetup{urlcolor=cyan}
	[\href{https://inspirehep.net/record/1427287}{inSPIRE}]}

  
\bibitem{Vilson.Some}
  Vilson O 2015 Some remarks concerning invariant quantities in scalar-tensor gravity {\it to appear in Advances in Applied Clifford Algebras} DOI:\href{http://dx.doi.org/10.1007/s00006-015-0567-4}{10.1007/s00006-015-0567-4}
  %
  {\hypersetup{urlcolor=violet}
  (arXiv:\href{https://arxiv.org/abs/1509.02481}{1509.02481}}) 
  %
  {\hypersetup{urlcolor=cyan}
  [\href{https://inspirehep.net/record/1392448}{inSPIRE}]}


\bibitem{JKV.2016}
  Kuusk P, J\"arv L and Vilson O 2016 Invariant quantities in the multiscalar-tensor theories of gravitation {\it Int.\ J.\ Mod.\ Phys.} A \href{http://dx.doi.org/10.1142/S0217751X16410037}{{\bf 31} 1641003}
  %
  {\hypersetup{urlcolor=violet}
  (arXiv:\href{https://arxiv.org/abs/1509.02903}{1509.02903})} 
  %
  {\hypersetup{urlcolor=cyan}
  [\href{https://inspirehep.net/record/1392641}{inSPIRE}]}
  

\bibitem{Damour.Nordtvedt}
  Damour T and Esposito-Far\`{e}se G 1992 Tensor-multi-scalar theories of gravitation {\it Class.\ Quantum Grav.} \href{http://dx.doi.org/10.1088/0264-9381/9/9/015}{{\bf 9} 2093--176}
  %
  {\hypersetup{urlcolor=cyan}
  [\href{http://inspirehep.net/record/332753}{inSPIRE}]}
  \\
  %
  %
  %
  Damour T and Nordtvedt K 1993 Tensor-scalar cosmological models and their relaxation toward general relativity {\it Phys.\ Rev.} D \href{http://dx.doi.org/10.1103/PhysRevD.48.3436}{{\bf 48} 3436--50}
  %
  {\hypersetup{urlcolor=cyan}
  [\href{http://inspirehep.net/record/356664}{inSPIRE}]}


\bibitem{LorenzPetzold} 
  Lorenz-Petzold D 1983 Comment on the general vacuum solutions in Brans-Dicke cosmology {\it Astrophys.\ Space Sci.} \href{http://dx.doi.org/10.1007/BF00651688}{{\bf 96} 451--3}


\bibitem{Domenech:2015qoa} 
  Dom\`{e}nech G and Sasaki M 2015 Conformal frame dependence of inflation {\it J.\ Cosmol.\ Astropart.\ Phys.} \href{http://dx.doi.org/10.1088/1475-7516/2015/04/022}{JCAP04(2015)022}
  %
  {\hypersetup{urlcolor=violet}
  	(arXiv:\href{https://arxiv.org/abs/1501.07699}{1501.07699})} 
  %
  {\hypersetup{urlcolor=cyan}
  	[\href{https://inspirehep.net/record/1342240}{inSPIRE}]}
  \\
  %
  %
  %
  Dom\`{e}nech G and Sasaki M 2016 Conformal frames in cosmology
  %
  {\hypersetup{urlcolor=violet}
  	arXiv:\href{https://arxiv.org/abs/1602.06332}{1602.06332}} 
  %
  {\hypersetup{urlcolor=cyan}
  	[\href{https://inspirehep.net/record/1422742}{inSPIRE}]}


\bibitem{Stewart:1993bc} 
  Stewart E D and Lyth D H 1993 A more accurate analytic calculation of the spectrum of cosmological perturbations produced during inflation {\it Phys.\ Lett.} B \href{http://dx.doi.org/10.1016/0370-2693(93)90379-V}{{\bf 302} 171--5}
  %
  {\hypersetup{urlcolor=violet}
	(arXiv:\href{https://arxiv.org/abs/gr-qc/9302019}{gr-qc/9302019})}
  %
  {\hypersetup{urlcolor=cyan}
	[\href{https://inspirehep.net/record/352726}{inSPIRE}]}


\bibitem{Mukhanov.Sasaki}
  Mukhanov V F 1985 Gravitational instability of the universe filled with a scalar field {\it JETP Letters} \href{http://www.jetpletters.ac.ru/ps/1467/article_22380.shtml}{{\bf 41} 493--6} ({\it Pis'ma Zh.\ Eksp.\ Teor.\ Fiz.} \href{http://www.jetpletters.ac.ru/ps/81/article_1454.shtml}{{\bf 41} 402--5})
  %
  {\hypersetup{urlcolor=cyan}
  	[\href{https://inspirehep.net/record/222017}{inSPIRE}]}
  \\
  %
  %
  %
  Sasaki M 1986 Large scale quantum fluctuations in the inflationary universe {\it Prog.\ Theor.\ Phys.} \href{http://dx.doi.org/10.1143/PTP.76.1036}{{\bf 76} 1036--46}
  %
  {\hypersetup{urlcolor=cyan}
  	[\href{https://inspirehep.net/record/230212}{inSPIRE}]}


\bibitem{Hwang:2005hb} 
  Hwang J-c and Noh H 2005 Classical evolution and quantum generation in generalized gravity theories including string corrections and tachyons: unified analyses {\it Phys.\ Rev.} D \href{http://dx.doi.org/10.1103/PhysRevD.71.063536}{{\bf 71} 063536}
  %
  {\hypersetup{urlcolor=violet}
  (arXiv:\href{https://arxiv.org/abs/gr-qc/0412126}{gr-qc/0412126})}
  %
  {\hypersetup{urlcolor=cyan}
  [\href{https://inspirehep.net/record/674212}{inSPIRE}]}


\bibitem{Gong:2011qe} 
  Gong J-O, Hwang J-c, Park W I, Sasaki M and Song Y-S 2011 Conformal invariance of curvature perturbation {\it J.\ Cosmol.\ Astropart.\ Phys.} \href{http://dx.doi.org/10.1088/1475-7516/2011/09/023}{JCAP09(2011)023}
  %
  {\hypersetup{urlcolor=violet}
  (arXiv:\href{https://arxiv.org/abs/1107.1840}{1107.1840})} 
  %
  {\hypersetup{urlcolor=cyan}
  [\href{https://inspirehep.net/record/917608}{inSPIRE}]}

  
\bibitem{JKSV.Parametrizations}
  J\"arv L, Kuusk P, Saal M and Vilson O 2014 Parametrizations in scalar-tensor theories of gravity and the limit of general relativity {\it J.\ Phys.:\ Conf.\ Ser.} \href{http://dx.doi.org/10.1088/1742-6596/532/1/012011}{{\bf 532} 012011}
  %
  {\hypersetup{urlcolor=violet}
  	(arXiv:\href{https://arxiv.org/abs/1501.07781}{1501.07781})} 
  %
  {\hypersetup{urlcolor=cyan}
  	[\href{http://inspirehep.net/record/1316183}{inSPIRE}]}
	

\bibitem{BDBW}
  Brans C and Dicke R H 1961 Mach's principle and a relativistic theory of gravitation {\it Phys.\ Rev.} \href{http://dx.doi.org/10.1103/PhysRev.124.925}{{\bf 124} 925--35}
  %
  {\hypersetup{urlcolor=cyan}
  	[\href{https://inspirehep.net/record/2450}{inSPIRE}]}
  \\
  %
  %
  %
  Bergmann P G 1968 Comments on the scalar-tensor theory {\it Int.\ J.\ Theor.\ Phys.} \href{http://dx.doi.org/10.1007/BF00668828}{{\bf 1} 25--36}
  %
  {\hypersetup{urlcolor=cyan}
  	[\href{https://inspirehep.net/record/52795}{inSPIRE}]}
  \\
  %
  %
  %
  Wagoner R V 1970 Scalar-tensor theory and gravitational waves {\it Phys.\ Rev.} D \href{http://dx.doi.org/10.1103/PhysRevD.1.3209}{{\bf 1} 3209--16}
  %
  {\hypersetup{urlcolor=cyan}
  	[\href{https://inspirehep.net/record/60983}{inSPIRE}]}


\bibitem{BEPS}
  Boisseau B, Esposito-Far\`ese G, Polarski D and Starobinsky A A 2000 Reconstruction of a scalar-tensor theory of gravity in an accelerating universe {\it Phys.\ Rev.\ Lett.} \href{http://dx.doi.org/10.1103/PhysRevLett.85.2236}{{\bf 85} 2236--9}
  %
  {\hypersetup{urlcolor=violet}
  	(arXiv:\href{https://arxiv.org/abs/gr-qc/0001066}{gr-qc/0001066})}
  %
  {\hypersetup{urlcolor=cyan}
  	[\href{https://inspirehep.net/record/523205}{inSPIRE}]}
  \\
  %
  %
  %
  Esposito-Far\`ese G and Polarski D 2001 Scalar-tensor gravity in an accelerating universe {\it Phys.\ Rev.} D \href{http://dx.doi.org/10.1103/PhysRevD.63.063504}{{\bf 63} 063504}
  %
  {\hypersetup{urlcolor=violet}
  	(arXiv:\href{https://arxiv.org/abs/gr-qc/0009034}{gr-qc/0009034})}
  %
  {\hypersetup{urlcolor=cyan}
  	[\href{https://inspirehep.net/record/533207}{inSPIRE}]}

\end{thebibliography}
\end{document}